\documentclass[12pt]{article} 
\pdfoutput=1
\usepackage{latexsym,epsfig,amssymb,amsmath}
\usepackage{color}
\usepackage{cancel}
\newcommand{\be}{\begin{equation}}
\newcommand{\ee}{\end{equation}}
\newcommand{\ben}{\begin{displaymath}}
\newcommand{\een}{\end{displaymath}}
\newcommand{\bea}{\begin{eqnarray}}
\newcommand{\eea}{\end{eqnarray}}

\newcommand{\bean}{\begin{eqnarray*}}
\newcommand{\eean}{\end{eqnarray*}}
\newcommand{\beqs}{\begin{eqnarray}}
\newcommand{\eeqs}{\end{eqnarray}}

\newcommand{\mathon}{\mathversion{bold}}
\newcommand{\mathoff}{\mathversion{normal}}

\makeatletter
\@addtoreset{equation}{section}
\makeatother

\begin{document}

\thispagestyle{empty}

\begin{flushright}\small
\end{flushright}


\bigskip
\bigskip

\mathon
\vskip 10mm

\begin{center}

  {\Large{\bf Matrix model holography}}

\end{center}

\mathoff


\vskip 6mm

\begin{center}
{\bf Thomas Ortiz$^a$, Henning Samtleben$^a$, and Dimitrios Tsimpis$^b$}\\[3ex]

$^a$\, {\em Universit\'e de Lyon, Laboratoire de Physique, UMR 5672, CNRS\\
\'Ecole Normale Sup\'erieure de Lyon\\
46, all\'ee d'Italie, F-69364 Lyon cedex 07, France}\\
\bigskip

$^b$\,{\em Universit\'{e} de Lyon\\
UMR 5822, CNRS/IN2P3, Institut de Physique Nucl\'{e}aire de Lyon\\ 
4 rue Enrico Fermi,  
F-69622 Villeurbanne Cedex, France}\\

\end{center}

\vskip1.8cm

\begin{center} {\bf Abstract } \end{center}

\begin{quotation}

\noindent
We set up the formalism of holographic renormalization for the matter-coupled
two-dimensional maximal supergravity that captures the low-lying fluctuations around the non-conformal 
D0-brane near-horizon geometry.
As an application we compute holographically one- and two-point functions of the BFSS matrix quantum mechanics and
its supersymmetric $SO(3)\times SO(6)$ deformation.

\end{quotation}

\newpage
\setcounter{page}{1}


\newpage

\section{Introduction}

`Matrix theory' or `matrix model', the theory of $\mathcal{N}=16$ supersymmetric $SU(N)$ gauged matrix quantum mechanics, was proposed in \cite{Banks:1996vh} as a nonperturbative formulation of M-theory. Genuine tests of the BFSS proposal, that is  tests which are not guaranteed to work solely by virtue of supersymmetric non-renormalization theorems, have been performed using Monte Carlo methods in a regime where the matrix quantum mechanics is strongly coupled. On the other hand the BFSS proposal can be understood within the framework of gauge/gravity duality: the holographic dual of matrix theory is a lightlike compactification of M-theory in an $SO(9)$-symmetric pp-wave background; moreover compactification to ten dimensions leads to an alternative interpretation whereby weakly-coupled IIA string theory in the near-horizon limit of $N$ $D0$ branes is the holographic dual of $SU(N)$ matrix  theory.

The gauge/gravity correspondence thus allows one to probe the strong-coupling limit of matrix theory using classical IIA supergravity in a 
conformal $AdS_2$ times $S^8$ background, which is the near-horizon geometry of D0 branes. This background 
can be thought of as the uplift to ten dimensions of a domain-wall solution of an effective two-dimensional dilaton-gravity theory. The latter theory is in fact a consistent truncation of IIA supergravity and can thus in principle be used to compute correlation functions in the matrix model involving the operators dual to the graviton and the dilaton, along the lines of holography for non-conformal branes \cite{Itzhaki:1998dd,Boonstra:1998mp,Kanitscheider:2008kd,Kanitscheider:2009as}. However since in two dimensions the dilaton and the graviton can both be gauged away at the classical level, one expects that the corresponding correlation functions  should be trivial;  we will see that this is indeed consistent with the results of the present paper. 

To go beyond trivial correlation functions one would need a two-dimensional consistent truncation of IIA which keeps more fields than just the metric and the dilaton. Although an effective lower-dimensional theory is not necessary for holography~\cite{Skenderis:2006uy},  it can help streamline the holographic computations along the lines of holographic renormalisation  \cite{Bianchi:2001de,Bianchi:2001kw,Skenderis:2002wp}. Recently a maximally-supersymmetric two-dimensional $SO(9)$ gauged supergravity was constructed in \cite{Ortiz:2012ib}. This theory is expected to be a consistent truncation of IIA supergravity on  $S^8$. Subsequently in \cite{Anabalon:2013zka} it was shown that a $U(1)^4$ truncation of the full $SO(9)$ gauge group is indeed a consistent truncation of IIA, and the uplift to ten dimensions was explicitly constructed. In particular the conformal  $AdS_2$ times $S^8$ near-horizon geometry was recovered as the uplift to ten dimensions of a supersymmetric domain-wall solution of the two-dimensional theory with sixteen supercharges.

In the present paper we will use the half-supersymmetric domain-wall solution of the  two-dimensional supergravity to compute correlation functions in 
the strongly-coupled matrix model using the prescription of holographic renormalization. In particular we compute two-point functions for the operators dual to scalars transforming in the ${\bf 44}$ and 
the ${\bf 84}$ of $SO(9)$.\footnote{The scalar sector of the two-dimensional maximally supersymmetric $SO(9)$ gauged supergravity contains, besides the dilaton, scalar fields 
transforming in the  ${\bf 44}\oplus{\bf 84}$ of  $SO(9)$; its $U(1)^4$ truncation contains the dilaton, four scalars coming from the ${\bf 44}$ and four scalars from the ${\bf 84}$ of $SO(9)$.} Our results 
are in agreement with the two-point functions previously computed both holographically, from the Kaluza-Klein spectrum of eleven-dimensional supergravity on $S^8$ \cite{Sekino:1999av,Sekino:2000mg}, and directly  in the matrix model by Monte Carlo methods \cite{Hanada:2011fq}.

Furthermore we  construct a half-supersymmetric `deformed' domain-wall solution of two-dimensional $SO(9)$  supergravity which uplifts to an eleven-dimensional pp-wave with symmetry broken from $SO(9)$ to $SO(3)\times SO(6)$. 
To achieve this deformation we must consider $SO(3)\times SO(6)$-preserving profiles for the scalar fields that
 go beyond the $U(1)^4$ truncation. As it turns out the resulting eleven-dimensional pp-wave is not of the form of the holographic dual to the BMN matrix model \cite{Berenstein:2002jq} which preserves $\mathcal{N}=32$ supersymmetry;\footnote{It is well-known that all pp-waves of eleven-dimensional  supergravity preserve at least sixteen supercharges. The maximally supersymmetric pp-wave \cite{KowalskiGlikman:1984wv} can be thought of as the Penrose limit of either the $AdS_7\times S^4$ or the $AdS_4\times S^7$ background \cite{figuroa}, while there are pp-waves with various possible fractions of supersymmetry 
between $\mathcal{N}=16$ and $\mathcal{N}=32$ \cite{Gauntlett:2002cs}.} nor does it belong to the class 
of  bubbling M-theory geometries  of \cite{Lin:2004nb}. Rather we will show that this $SO(3)\times SO(6)$ deformation should be identified holographically with a vev deformation of the BFSS matrix model. 

As in the undeformed case we use holographic renormalization to compute two-point correlation functions of operators dual to the scalar fields in the  ${\bf 44}$ of  $SO(9)$. More precisely, under $SO(3)\times SO(6)$ the ${\bf 44}$ decomposes as ${\bf(1,1)\oplus(1,20)\oplus(5,1)\oplus(3,6)}$; the three distinct two-point functions that we compute in the present paper are those of operators dual to the scalars  
outside the ${\bf (1,1)}$ singlet.\footnote{This choice was made for simplicity, since the singlet would mix already 
at the quadratic level with the operators 
coming from the other representations.} We have checked numerically that 
in the UV-limit all three reduce to the two-point function of the ${\bf 44}$ scalar computed in the undeformed matrix model. 
This is consistent with the fact that the deformed domain-wall solution reduces in the limit of small radial direction to the undeformed 
domain wall. Equivalently it can be checked that the ten-dimensional uplift of the deformed domain-wall solution is asymptotically conformal $AdS_2$ times $S^8$.

The plan of the remainder of the paper is as follows.  Section~\ref{sec:BFSS} discusses holographic renormalization for
the two-dimensional maximal $SO(9)$ supergravity dual to the BFSS matrix quantum mechanics.
As a warm-up we compute one- and two-point functions for the operators dual to the graviton and the dilaton and show that they are 
trivial as expected. We then extend the computation to one- and two-point functions in the scalar sector, where we reproduce the
expected field theory results for the corresponding operators. 
In section \ref{sec:deformation} we construct a half-supersymmetric domain-wall solution of supergravity which breaks
$SO(9)$ down to $SO(3)\times SO(6)$ and is expected to provide a holographic description of a corresponding vev deformation 
of the matrix model. We set up the holographic renormalization around this background and in particular 
 compute the deformed correlation functions in the scalar sector.
 Some future directions are discussed in section~\ref{sec:discussion}.
In appendix \ref{app:validity} we review the various holographic dualities 
of the matrix model and their respective regimes of validity.
In appendix~\ref{app:amb} we review the ambiguity in the holographic dictionary 
for scalar fields in a certain mass range which will be relevant for our model.

\section{BFSS and holographic renormalization}
\label{sec:BFSS}

In this section, we will employ the effective two-dimensional supergravity that describes fluctuations
around the D0-brane near-horizon geometry, and apply the procedure of holographic renormalization in order
to extract one- and two-point correlation functions of the corresponding operators in the dual matrix quantum mechanics.

\subsection{Effective 2d supergravity and fluctuation equations}

The two-dimensional maximally supersymmetric $SO(9)$ supergravity 
constructed in~\cite{Ortiz:2012ib} 
 describes fluctuations
around the $S^8$ compactification of IIA supergravity. The full theory carries a dilaton $\rho$ and 128 scalar fields, 
transforming as ${\bf 44\oplus 84}$ under  $SO(9)$\,.
Here, we will only consider its $U(1)^4$ truncation which apart from 
$\rho$ and the $U(1)^4$ gauge fields carries four more dilaton fields $u_{a}$ from the ${\bf 44}$
and four axion fields $\phi_{a}$ from the ${\bf 84}$ of $SO(9)$\,. 
The truncated action is given by~\cite{Anabalon:2013zka}
\bea
 {\cal{L}}&= &  -\frac{1}{4} e \rho \, R + \frac{1}{2} e \rho \sum_{a} \partial_{\mu} u_{a} \,  \partial^{\mu} u_{a}
 + \frac{1}{2} e \rho^{1/3} X_{0}^{-1} \sum_{a=1}^{4} X_{a}^{-2} \left( \partial_{\mu} \phi^{a} \right) \left( \partial^{\mu} \phi^{a} \right)
\nonumber\\
&&{}- \frac{\rho}{8} \, \varepsilon^{\mu \nu} F_{\mu\nu}^{a} \; y^{a}   - e \,V_{\text{pot}}\;,
\label{Ltrunc}
\eea
where we have defined $X_{0} \equiv \prod_a X_a^{-2} $, the scalar kinetic term is defined via
\bea
X_{a} \equiv  e^{-2\,A_{ab} u_{b}}\;,\qquad
A ~\equiv~ {\footnotesize \begin{pmatrix}
  1/6 & -1/\sqrt{2} & -1/\sqrt{6} & -1/(2\sqrt{3})\\
  1/6 & 0 & 0 & \sqrt{3}/2\\
  1/6 & 0 & \sqrt{2/3} & -1/(2\sqrt{3})\\
  1/6 & 1/\sqrt{2} & -1/\sqrt{6} &  -1/(2\sqrt{3})\\
 \end{pmatrix}}
 \;,
\eea
and the abelian field strengths
$F_{\mu\nu}^{a} \equiv 2 \, \partial^{\vphantom{a}}_{[\mu} A_{\nu]}^{a}$ couple to 
four auxiliary scalar fields~$y^a$ that can be integrated out from the action.
The scalar potential of (\ref{Ltrunc}) is given by
\bea
V_{\text{pot}} &=&  \rho^{5/9} \,\Big[ \frac{1}{8} \,\Big({X_{0}}^{2} - 8 \sum_{a<b} X_{a} X_{b} - 4 X_{0} \sum_{a} X_{a} \Big)
+ \frac{1}{2} \, \rho^{-2/3}  \sum_{a} X_{a}^{-2}  \left( X_{0} - 4 X_{a} \right) (\phi^{a})^{2} \nonumber\\
&&\qquad\quad{}  + 2 \, \rho^{-4/3} \sum_{a<b}X_{a}^{-2} X_{b}^{-2} (\phi^{a})^{2} (\phi^{b})^{2}
+ \frac18 \, \rho^{-2} \sum_{a} X_{a} \,  \Big(\rho \, y^{a} + 8 \,  \prod_{b \neq a} \phi^{b} \Big)^{2} \nonumber\\
&&\qquad\quad{}  + \frac12 \, \rho^{-8/3} X_{0}^{-1} \Big( \sum_{a}  \, \rho \,y^{a} \phi^{a} + 8  \prod_{a} \phi^{a} \Big)^{2}  \, \Big]
\;,
\eea
as a fourth order polynomial in the scalars $\phi^a$\,.
The action (\ref{Ltrunc}) admits a half supersymmetric domain wall solution, in which all scalars and gauge fields
vanish and metric and dilaton are given by
\bea
\label{dw}
ds^2 &=&  r^7 dt^2 - dr^2\;,\qquad
\rho(r)~=~ r^{9/2}\;.
\eea
This two-dimensional solution can be uplifted into type IIA supergravity as
\bea
 ds_{10}^2 = r^{-7/8} \big( r^7 \text{dt}^2 - (\text{dr}^2 + r^2 \, d\Omega_{8}^2)\big)
 \;,\quad
\Phi = -\frac{21}{8} \ln r\;,\quad
F = d\, \big( r^7  \text{dt} \big)  \;,
\label{abh}
\eea
(with 10D dilaton $\Phi$ and two-form flux $F$)
and further to an eleven-dimensional pp-wave solution
\cite{Hull:1984vh,Townsend:1998qp,Gauntlett:2002cs}
\begin{equation}\label{2.6}
ds_{11}^{2}= dx^+ \, dx^- + (1-r^{-7}) (dx^-)^2 - (dr^2+ r^2 \,d\Omega_{8}^{2}) \,.
\end{equation}

In this section, we will compute correlation functions associated 
to the quadratic fluctuations around the domain wall (\ref{dw}).
Since scalars originating from different $SO(9)$ representations do not mix at the quadratic level, 
we will only need the truncated action (\ref{Ltrunc}) of two-dimensional dilaton gravity coupled to one of the scalars $X_a$ and one of the scalars $\phi^a$. We will denote these two scalars by $y_{44}$ and $y_{84}$ respectively (referring to their $SO(9)$ origin), 
and collectively by $y_n$.  Moreover, it will be convenient to go to a frame in which 
the background metric of (\ref{dw}) becomes pure AdS which is 
achieved by rescaling the fields as
\begin{equation}
t\rightarrow \frac{2}{5} \, t \,,\quad r \rightarrow \,r^{-1/5}\,, \quad
g_{\mu \nu} \rightarrow \frac{4}{25} \,\rho^{4/9} \, g_{\mu \nu}\,.
\end{equation}
In this frame, and after Wick rotation to Euclidean signature, 
the action takes the canonical form~\cite{Kanitscheider:2008kd}
\begin{equation}
\label{effy}
S = \frac{1}{4} \, \int d^{2}x\, \sqrt{|g|} \, e^{\gamma \phi} \left( R + \beta  \left(\partial \phi \right)^{2} + C 
- e^{a_n\phi} \left( \left( \partial y_n \right)^{2} - m_n^2 \, y_n^{2} \right) \right) .
\end{equation}
with $\rho\equiv e^{\gamma \phi}$, and the constants
\begin{equation}
\gamma \equiv  -\frac{6}{7} \,,\quad  \beta\equiv\frac{16}{49} \,,\quad  C\equiv\frac{126}{25}\,,
\end{equation}
describing the dilaton-gravity sector. With these coordinates, the boundary of AdS is located at ${r=0}$
and the background (\ref{dw}) takes the form
\begin{equation}
\label{backG}
ds^2 = \frac{1}{r} dt^2 + \frac{1}{4r^2}dr^2\,,\qquad
e^\phi = r^{\alpha} \,,\qquad \alpha\equiv \frac{21}{20}\;.
\end{equation}
The scalar couplings in (\ref{effy}) are characterized by the constants $a_n$ and $m_n$ which take different
values for the scalars in the 44 and 84, respectively:
\bea
\label{amy}
&&{} a_{44}\equiv0 \,, \quad m_{44}^2\equiv\frac{8}{5}\,,\;\; \quad y_{44} \equiv 6\sqrt{2} \,x \,,
\quad\mbox{with}\quad 
X_{1,2,3,4}=e^{-2x}\;,
\nonumber\\
&&{} a_{84}\equiv \frac{4}{7} \,, \quad m_{84}^2\equiv\frac{12}{25}\,, \quad y_{84} \equiv \sqrt{2} \,\phi^{a=1} \,.
\eea
Let us note that the addition of scalar matter in (2.8) is the source of some technical
complications with respect to the standard treatment of the dilaton gravity 
sector~\cite{Kanitscheider:2008kd,Kanitscheider:2009as}. 
In particular the fact that the scalars $y_{84}$ arise with a non-vanishing relative
dilaton power $a_{84}$ prevents us from using the methods of~\cite{Kanitscheider:2009as}
and translate the non-conformal holographic problem into a pure AdS background in some suitable higher dimension. 
However, it is straightforward to extend the analysis of \cite{Kanitscheider:2008kd} to the presence of additional
matter fields.

The equations of motion follow from (\ref{effy}) and yield
\begin{align}
\label{eomy}
0&= \left( \nabla_{\mu} \partial_{\nu} \phi \right)  -\frac{g_{\mu \nu}}{2}  \nabla \partial \phi 
- \Big( \frac{\beta}{\gamma} - \gamma \Big)
 \Big( \left( \partial_{\mu} \phi \right) \left( \partial_{\nu} \phi \right) - \frac{g_{\mu \nu}}{2}  \left( \partial \phi \right)^2  \Big) \nonumber\\
&\quad+\frac{e^{a_n\phi}}{\gamma} \big( \partial_{\mu} y_n \partial_{\nu} y_n 
- \frac{1}{2} g_{\mu \nu} (\partial y_n)^2 \big)\,,\nonumber\\  
0&= \gamma \nabla \partial \phi + \gamma^2 \big( \partial \phi \big)^2 - C - m_n^2 e^{a_n\phi} y_n^2\,,\nonumber\\
0&=  R - 2\frac{\beta}{\gamma} \nabla \partial \phi  - \beta \left( \partial \phi \right)^{2} + C 
- \big(1+\frac{a_n}{\gamma} \big) e^{a_n\phi} 
\left( \left( \partial y \right)^{2} - m_n^2 \, y_n^{2} \right)  \,,\nonumber\\
0&= \nabla^{\mu} \big( e^{(a_n+\gamma)\phi}\, \partial_{\mu} y_n \big) + m_n^2 e^{(a_n+\gamma)\phi}\, y_n\,.
\end{align}
They respectively stand for: the traceless and trace part of Einstein equations, the dilaton field equation, and
the scalar equations of motion.


\subsection{Asymptotic expansions}

Following the procedure of holographic renormalization \cite{Bianchi:2001de,
Bianchi:2001kw,Skenderis:2002wp,Kanitscheider:2008kd},
we first compute the asymptotic expansions of all fields at the boundary $r=0$. 
As an illustration, let us first restrict to the dilaton-gravity sector, i.e.\ set all scalar fields other than the dilaton to zero, in which case we
reproduce the results of \cite{Kanitscheider:2008kd} for the (degenerate) case of the D0 branes.
The fluctuation ansatz for metric and dilaton is given by
\begin{equation}
\begin{aligned}
ds^2 &= \frac{f(t,r)}{r} dt^2 + \frac{1}{4r^2}dr^2\,,\\
\phi &= \alpha\, \ln r + \frac{\kappa(t,r)}{\gamma}\,.\\
\end{aligned}
\end{equation}
with functions $f(t,r)$, $\kappa(t,r)$ admitting a (fractional) power expansion in $r$ near $r=0$
\bea
f(t,r)= f_{(0)}(t) + \underset{r \to 0}{o}(1) \,,\qquad
\kappa(t,r)= \kappa_{(0)}(t) + \underset{r \to 0}{o}(1) \,.
\eea
According to the equations of motion \eqref{eomy}, the functions $f(t,r)$ and $\kappa(t,r)$ are subject to 
the non-linear partial differential equations
\begin{align}
\label{efk}
0&= - \frac{1}{4} \big( f^{-1} f' \big)^{2} + \frac{1}{2} f^{-1} f'' + \kappa'' + \big( 1-\frac{\beta}{\gamma^{2}}\big) \big( \kappa'\big)^{2} \,, \nonumber\\
0&=\big(1-\frac{\beta}{\gamma^{2}} \big) \dot{\kappa} \kappa' + \dot{\kappa}' - \frac{1}{2} f' f^{-1} \dot{\kappa} \,,\\
0&=2 \alpha \gamma f' + r \big( 2 f'' - f^{-1} \big( f' \big)^{2}\big) + \ddot{\kappa} - \frac{1}{2} f^{-1} \dot{f} \dot{\kappa}
+ \big(1-\frac{\beta}{\gamma^{2}} \big)\big( \dot{\kappa}\big)^{2} - 2f \big( 1 - r f^{-1} f'\big) \kappa' \,,\nonumber\\
0&=4r \big( \kappa'' + \big( \kappa' \big)^{2} \big) + \big( 8 \alpha \gamma + 2 + 2rf^{-1} f' \big) \kappa'+
f^{-1} \big( \ddot{\kappa}  - \frac{1}{2} f^{-1} \dot{f} \dot{\kappa}+ \big( \dot{\kappa}\big)^{2}  \big) + 2 f^{-1} f' \alpha \gamma \,,\nonumber
\end{align}
where dots and primes refer to $\partial_t $ and $\partial_r$, respectively. 
Closer inspection of these equations shows that its solutions admit a fractional power expansion around $r=0$
\begin{align}
f(t,r)&=f_{(0)}(t) + r \, f_{(5)}(t)+  r^{\sigma} \,  f_{(5\sigma)}(t) + \dots  \,,\nonumber\\
\kappa(t,r)&=\kappa_{(0)}(t) + r \, \kappa_{(5)}(t)+  r^{\sigma} \,  \kappa_{(5\sigma)}(t) + \dots  \,,
\label{fk0}
\end{align}
where $\sigma = \frac{1}{2} - \alpha \gamma=\frac{7}{5}$ denotes the first non-integer power 
in the expansion, whose coefficient is not determined by the equations of motion (\ref{efk}).
In generic dimensions, this coefficient carries the information about the two-point correlation functions
of the associated operators. In two dimensions (i.e.\ for the $p=0$ branes) this structure
is highly degenerate.
Specifically, the equations of motion (\ref{efk}) determine the coefficients 
$\kappa_{(5)}$, $f_{(5)}$ as
\begin{align}
\label{f5}
  \kappa_{(5)}&=\frac{5}{36} \, f_{(0)}^{-1} {\dot{\kappa}}_{(0)}^{2} \,,\nonumber\\
   f_{(5)} &= \frac{5}{9} \big( {\ddot{\kappa}}_{(0)} 
	- \frac{1}{2} f_{(0)}^{-1} {\dot{f}}_{(0)} {\dot{\kappa}}_{(0)}
	+ \frac{5}{18} {\dot{\kappa}}_{(0)}^{2}   \big) \,,
 \end{align}
and constrain the coefficients $\kappa_{(5\sigma)}$, $f_{(5\sigma)}$ as
\begin{align}
\label{fs}
0&=f_{(5\sigma)} + 2 f_{(0)} \kappa_{(5\sigma)} \,,\nonumber\\
0&=\dot{\kappa}_{(5\sigma)} + \frac{14}{9}  \dot{\kappa}_{(0)} \kappa_{(5\sigma)} \,.
\end{align}
The latter conditions imply the two-dimensional analogue of what in higher dimensions 
expresses the diffeomorphism and trace Ward identities \cite{Bianchi:2001kw,Kanitscheider:2008kd}.
In two dimensions these contraints imply that there are no non-trivial correlation functions
associated to the operators dual to $f$ and $\kappa$, respectively, as we shall discuss shortly.
This is related to the fact that in two dimensions the dilaton-gravity sector does not carry any propagating
degrees of freedom. In this case, the interesting structure is sitting in the scalar sector of the theory.
Let us thus repeat the previous analysis in presence of the scalar fields.

Consider first the action (\ref{effy}) with scalar fields from the ${\bf 44}$ and the ${\bf 84}$ of $SO(9)$. 
The equations of motion obtained from variation of (\ref{effy}) then imply a generalization of 
the ansatz (\ref{fk0}) to a fractional expansion of the type
\begin{align}
\label{ax}
f(t,r)&=f_{(0)}(t) + r^{4/5} \, f_{(4)}(t)+  r \, f_{(5)}(t) + r^{7/5} \,f_{(7)}(t) + \dots\;,\nonumber\\
\kappa(t,r)&=\kappa_{(0)}(t) + r^{4/5} \, \kappa_{(4)}(t)
+  r \, \kappa_{(5)}(t) + r^{7/5} \, \kappa_{(7)}(t) + \dots\;,\nonumber\\
y_{44}(r,t)&= r^{2/5} \, x_{(2)}(t) +  r \, x_{(5)}(t)  + \dots \;,\nonumber\\
y_{84}(r,t)&= r^{1/5} \, y_{(1)}(t) +  r^{3/5} \, y_{(3)}(t)  + \dots \;,
\end{align}
where $x_{(5)}$ and $y_{(3)}$ correspond to the coefficients in the scalar expansion 
that are left undetermined by the equations of motion. The intermediate coefficients in the series expansion
are determined by the equations of motion to
\begin{align}
\label{cx}
\kappa_{(4)} &= - \frac{1}{4} \, x_{(2)}^{2} \,,\nonumber\\
\kappa_{(5)} &=\frac{5}{36} \, f_{(0)}^{-1} {\dot{\kappa}}_{(0)}^{2}  - \frac{1}{10} e^{-\frac{2 \kappa_{(0)}}{3}} y_{(1)}^{2} 
\,,\nonumber\\
\dot{\kappa}_{(7)} &= - \frac{14}{9} \dot{\kappa}_{(0)} \kappa_{(7)}
-\frac{e^{-\frac{2}{3}\kappa_{(0)}}}{7} \big(3 \dot{y}_{(1)} y_{(3)} + y_{(1)} \dot{y}_{(3)}
+ \frac{4}{3} y_{(1)} y_{(3)} \dot{\kappa}_{(0)} \big) \nonumber\\
&\quad - \frac{1}{7} \big( 5 \dot{x}_{(2)} x_{(5)} + 2 x_{(2)} \dot{x}_{(5)} + \frac{40}{9} x_{(2)} x_{(5)} \dot{\kappa}_{(0)} \big) \,,\nonumber\\
 f_{(4)} &= - \frac{5}{18} f_{(0)} \, x_{(2)}^{2}\,,\nonumber\\
 f_{(5)} &= \frac{5}{9} \big( {\ddot{\kappa}}_{(0)}
 - \frac{1}{2} f_{(0)}^{-1} {\dot{f}}_{(0)} {\dot{\kappa}}_{(0)} + \frac{5}{18} {\dot{\kappa}}_{(0)}^{2}   \big)
 + \frac{1}{45} e^{-\frac{2 \kappa_{(0)}}{3}} f_{(0)} y_{(1)}^{2}\,, \nonumber\\
f_{(7)} &= -2 f_{(0)} \kappa_{(7)} - \frac{80}{63} f_{(0)} \, x_{(2)} x_{(5)}
- \frac{8}{21} e^{-\frac{2 \kappa_{(0)}}{3}} f_{(0)} y_{(1)} y_{(3)}\,.
\end{align}
In absence of the scalar fields these expressions consistently reproduce (\ref{f5}).

\subsection{Regularization and counterterms}

\paragraph{On-shell action}

The central object for the computation of correlation functions is the action (\ref{effy}) evaluated on-shell.
Using the dilaton field equation from (\ref{eomy}), the on-shell Lagrangian reduces to
\bea
{\cal L}|_{\rm on-shell} &=& 
\frac{2\beta}{\gamma} \sqrt{|\mathrm{det}g|} \,\nabla \big( e^{\gamma \phi} \partial \phi \big) 
+\frac{a_{84}}{\gamma} \sqrt{|\mathrm{det}g|} \,e^{a_{84} \phi} \left((\partial y_{84})^2-m^2 y_{84}^2\right)\,.
\label{Lonshell1}
\eea
Note that no explicit scalar dependence on $y_{44}$ appears in the Lagrangian. This is due to the fact
that these scalars appear coupled with the same dilaton power as the Einstein-Hilbert term, c.f.~(\ref{effy}), (\ref{amy}),
thus disappear form the action upon using the dilaton equation of motion.
Moreover, we need to add the Gibbons-Hawking term 
in order to take into account the boundary of the background spacetime 
\begin{equation}
\int_{{\cal M}} d^{2}x \, \sqrt{|\mathrm{det}g|} \,e^{\gamma \phi}\, R \quad \longrightarrow \quad
\int_{{\cal M}} d^{2}x \, \sqrt{|\mathrm{det}g|} \,e^{\gamma \phi}\, R
+ \int_{\partial {\cal{M}}} ds \sqrt{h} \,e^{\gamma \phi}\, 2\, K \,.
\end{equation}
Here $h$ is the induced metric on the (one-dimensional) boundary and $K$ is the trace of the extrinsic curvature
 of the boundary that can be computed from a unit length vector $n^{\mu}$ normal to the boundary
\begin{equation}
K = \nabla_{\mu} n^{\mu}\,.
\end{equation}
Putting everything together, the full on shell action is given by
\begin{equation}
\label{AdSEdW0}
S_{\text{on-shell}} = \frac{1}{2} \int_{\partial {\cal M}} dt
 \sqrt{h} \, e^{\gamma \phi} \left( K + \frac{\beta}{\gamma} n^{\mu} \partial_{\mu} \phi + \frac{2}{7\gamma} e^{\frac{4}{7} \phi} \, y_{84} \; n^{\mu} \partial_{\mu} y_{84}  \right)\,,
\end{equation}
where the boundary is located at $r=0$. Because the integral diverges when $r \to 0$, 
the first step of holographic renormalization consists in regularizing the integral by
introducing a parameter $\epsilon$ in order to control the divergences
\begin{equation}
\label{AdSEdW}
S_{\text{reg}} = \frac{1}{2} \int_{\partial {AAdS}, r=\epsilon} dt
 \sqrt{h} \, e^{\gamma \phi} \left( K + \frac{\beta}{\gamma} n^{\mu} \partial_{\mu} \phi + \frac{2}{7\gamma} e^{\frac{4}{7} \phi} \, y_{84} \; n^{\mu} \partial_{\mu} y_{84}  \right)\,.
\end{equation}
Knowing the asymptotic behaviour of the fields near the boundary, the regularized on-shell 
action \eqref{AdSEdW} may be evaluated as a function of $\epsilon$. 
Let us recall that $n^{\mu}$ is a unit vector ($n^{\mu} n_{\mu} = 1$) normal to the 
boundary
\begin{equation}
 n^{\mu} \partial_{\mu}= n \, \partial_r = 2r\partial_r \,,
\end{equation}
and
\bea
h = \frac{f(t,r)}{r} dt^{2}\,,\qquad
K = \nabla_{\mu} n^{\mu} = -1 + r \, \partial_{r} \ln f \,.
\eea
Inserting the expansion \eqref{ax} in the action \eqref{AdSEdW} leads to the different contributions
\begin{align}
\sqrt{h} \, e^{\gamma \phi}  &= |f_{(0)}|^{1/2} e^{\kappa_{(0)}} \, \epsilon^{-7/5}\, \Big[ 1 +
\big(\frac{1}{2} f_{(0)}^{-1} f_{(4)} + \kappa_{(4)} \big) \, \epsilon^{4/5}
+ \big(\frac{1}{2} f_{(0)}^{-1} f_{(5)} + \kappa_{(5)} \big) \, \epsilon \nonumber\\
&\hspace*{4cm}+ \big( \frac{1}{2} f_{(0)}^{-1} f_{(7)} + \kappa_{(7)} \big) \, \epsilon^{7/5}   \Big]+\dots \,,\nonumber\\
 \left.K\right|_{r=\epsilon}&= -1 + f_{(0)}^{-1} \, \Big[ \frac{4}{5} f_{(4)} \, \epsilon^{4/5}
 + f_{(5)} \, \epsilon + \frac{7}{5} f_{(7)} \, \epsilon^{7/5}  \Big] +\dots\,,\nonumber\\
 \left.n^{\mu} \partial_{\mu} \phi\right|_{r=\epsilon} &= 2\alpha + \frac{2}{\gamma} \Big[
\frac{4}{5} \kappa_{(4)} \, \epsilon^{4/5}
+ \kappa_{(5)} \epsilon + \frac{7}{5} \kappa_{(7)} \epsilon^{7/5} \Big]  +\dots \,,\nonumber\\
 \left.e^{\frac{4}{7} \phi} \, y \; n^{\mu} \partial_{\mu} y\right|_{r=\epsilon}   &= 
 e^{-\frac{2}{3} \kappa_{(0)}}  \, \Big[ \frac{2}{5} \, y_{(1)}^2 \, \epsilon + \frac{4}{5} \, y_{(1)} y_{(3)} \, 
\epsilon^{7/5} \Big] + \dots
\,.
\end{align}
The most divergent term in this expansion comes from the determinant of the induced metric times the dilaton and involves 
a global factor of $\epsilon^{-7/5}$. 
The on-shell action can now be expressed as a perturbative expansion in $r=\epsilon$ up to terms vanishing when $\epsilon$ goes to zero 
\bea
\label{Sdiv}
S_{\text{reg}} &=& \frac{1}{2} \int  dt \; |f_{(0)}|^{1/2} e^{\kappa_{(0)}} 
\left( L_{(-7)}\,\epsilon^{-7/5}+
L_{(-3)}\,\epsilon^{-3/5}+L_{(-2)}\,\epsilon^{-2/5}+L_{(0)}\,\epsilon^{0} + o(1)  \right)
\;,
\nonumber\\[2ex]
&&{} L_{(-7)} ~\equiv~ -1 + \frac{2\alpha \beta}{\gamma}~=~ -\frac95 \;,\\
&&{} L_{(-3)} ~\equiv~ -\frac95 \left( \frac{1}{2} f_{(0)}^{-1} f_{(4)} + \kappa_{(4)} \right)
+ \frac{4}{5} f_{(0)}^{-1} f_{(4)} + \frac{4}{5} \frac{2\beta}{\gamma^{2}} \kappa_{(4)} \;,\nonumber\\
&&{} L_{(-2)} ~\equiv~  -\frac95 \left(
\frac{1}{2} f_{(0)}^{-1} f_{(5)} + \kappa_{(5)} \right)
 + f_{(0)}^{-1} f_{(5)} + \frac{2\beta}{\gamma^{2}} \kappa_{(5)} + \frac{4}{35 \gamma} e^{-\frac{2}{3} \kappa_{(0)}} y_{(1)}^2  \;,\nonumber \\
&&{} L_{(0)} ~\equiv~ -\frac95 \left(
\frac{1}{2} f_{(0)}^{-1} f_{(7)} +  \kappa_{(7)} \right)
+ \frac{7}{5} f_{(0)}^{-1} f_{(7)} + \frac{7}{5}\frac{2\beta}{\gamma^{2}} \, \kappa_{(7)} 
+ \frac{16}{35 \gamma} e^{-\frac{2}{3} \kappa_{(0)}} y_{(1)} y_{(3)}\;.
\nonumber
\eea
We note that there is no explicit dependence on the scalars $x_{(2)}$, $x_{(5)}$, c.f.\ the discussion after (\ref{Lonshell1}).
The dependence of the regularized action on these fields enters implicitly via the metric and dilaton components (\ref{cx}).

\paragraph{Counterterms}

The first counter-term required for cancelling the most divergent contribution in (\ref{Sdiv}) 
takes the form of an exponential dilaton potential
\begin{equation}
S_{\text{ct1}} = \frac{1}{2} \int \text{dt}  \sqrt{h} \, e^{\gamma \phi} \Big(  1 - \frac{2\alpha \beta}{\gamma} \Big)\,.
\label{sct1}
\end{equation}
This kills the first divergent term in \eqref{Sdiv} 
and also modifies the sub-leading terms
\begin{equation}
\label{div1}
\begin{aligned}
  S_{\text{reg}} + S_{\text{ct1}} &= \frac{1}{2} \int dt \; |f_{(0)}|^{1/2} e^{\kappa_{(0)}} \Big[ 
\frac{4}{5} \big( f_{(0)}^{-1} f_{(4)} + \frac{2\beta}{\gamma^{2}} \kappa_{(4)} \big) \;\epsilon^{-3/5} \\
&\quad+\big( f_{(0)}^{-1} f_{(5)} + \frac{2\beta}{\gamma^{2}} \kappa_{(5)} + \frac{4}{35 \gamma} e^{-\frac{2}{3} \kappa_{(0)}} y_{(1)}^2 \big) \;\epsilon^{-2/5} \\
&\quad+ \frac{7}{5} \big( f_{(0)}^{-1} f_{(7)} + \frac{2\beta}{\gamma^{2}} \kappa_{(7)} \big) 
+\frac{16}{35 \gamma} e^{-\frac{2}{3} \kappa_{(0)}} y_{(1)} y_{(3)} \; +\; o(1)\Big]\,.\\
\end{aligned}
\end{equation}
Moreover, $f_{(5)}$ and $\kappa_{(5)}$ are related to the sources by \eqref{cx}. This corresponds to the expansion of
\begin{equation}
\begin{aligned}
\left.\big( \nabla^{t} \partial_{t} \phi \big)\right|_{r=\epsilon} &= \frac{f_{(0)}^{-1}}{\gamma} 
\big( {\ddot{\kappa}}_{(0)} - \frac{1}{2} f_{(0)}^{-1} {\dot{f}}_{(0)} {\dot{\kappa}}_{(0)} \big) \, \epsilon  + o(\epsilon)\,,\\
\left.\big( \partial \phi \big)^{2}\right|_{r=\epsilon} &= \frac{f_{(0)}^{-1} {\dot{\kappa}}_{(0)}^{2}}{\gamma^{2}} \, \epsilon  + o(\epsilon)\,,\\
\end{aligned}
\end{equation}
and determines the form of the second counter-term 
\begin{equation}
\begin{aligned}
 S_{\text{ct2}} &= \frac{1}{2} \int dt  \sqrt{h}
 \, e^{\gamma \phi} \Big( \frac{10}{21} \big( \nabla^{t} \partial_{t} \phi \big) 
- \frac{10}{49} \big( \partial \phi \big)^{2} \Big)\\
&= \frac{1}{2} \int dt |f_{(0)}|^{1/2} e^{\kappa_{(0)}} 
\big(-\frac{5}{9} f_{(0)}^{-1}\big)\Big(  {\ddot{\kappa}}_{(0)}
 - \frac{1}{2} f_{(0)}^{-1} {\dot{f}}_{(0)} {\dot{\kappa}}_{(0)} 
+ \frac{1}{2} {\dot{\kappa}}_{(0)}^{2} \Big) \, \epsilon^{-2/5} + o(1)\;,\\
\end{aligned}
\end{equation}
These terms cancel the $f_{(5)}$ and $\kappa_{(5)}$ contributions to the divergent part of the on-shell action (\ref{div1}).
Upon furthermore replacing $f_{(4)}$ and $\kappa_{(4)}$ 
by their expression from \eqref{cx}, the resulting action reads
\begin{equation}
\begin{aligned}
  S_{\text{reg}} + S_{\text{ct1}} + S_{\text{ct2}} &= \frac{1}{2} \int dt \; |f_{(0)}|^{1/2} e^{\kappa_{(0)}} \Big[ 
-\frac{2}{5}  x_{(2)}^2  \;\epsilon^{-3/5}
 - \frac{1}{5} e^{-\frac{2}{3} \kappa_{(0)}} y_{(1)}^2  \;\epsilon^{-2/5} \\
&\qquad\quad+ \frac{7}{5} \big( f_{(0)}^{-1} f_{(7)} + \frac{2\beta}{\gamma^{2}} \kappa_{(7)} \big) 
+\frac{16}{35 \gamma} e^{-\frac{2}{3} \kappa_{(0)}} y_{(1)} y_{(3)} \; +\; o(1)\Big]\,.\\
\end{aligned}
\end{equation}
From this expression we read off the last counterterms for the matter couplings
\bea
S_{\text{ct3}} &=& \frac{1}{5} \int dt  \sqrt{h} \, e^{\gamma \phi}\, y_{(44)}^{2}  \,,\nonumber\\
 S_{\text{ct4}} &=& \frac{1}{10} \int dt \sqrt{h} \, e^{(\gamma + a) \phi} \, y_{(84)}^{2}  \,.
 \eea
After renormalization by all counter-terms, the on-shell action is given by
\bea
S_{\text{ren}} &=& S_{\text{reg}} + S_{\text{ct1}} + S_{\text{ct2}}+ S_{\text{ct3}} + S_{\text{ct4}}
\\
 &=& \frac{1}{2} \int dt |f_{(0)}|^{1/2} e^{\kappa_{(0)}}
 \bigg[ \frac{7}{5} \big( f_{(0)}^{-1} f_{(7)} + \frac{2\beta}{\gamma^{2}} \kappa_{(7)} \big) 
 +\frac{4}{5} x_{(2)} x_{(5)}  - \frac{2}{15} e^{-\frac{2 \kappa_{(0)}}{3}} y_{(1)} y_{(3)}
 \bigg] 
\;.
\nonumber
\eea
and contains only finite terms in the limit $\epsilon\rightarrow0$. 
Eventually, taking into account the  relation between $f_{(7)}$ and $\kappa_{(7)}$ from \eqref{fs}, the renormalized
action takes the final form
\bea
\label{Sren}
 S_{\text{ren}} 
& =& 
\int \text{dt}|f_{(0)}|^{1/2} e^{\kappa_{(0)}} \left( - \frac{7}{9} \kappa_{(7)} 
- \frac{22}{45} x_{(2)} x_{(5)}  - \frac{1}{3} e^{-\frac{2 \kappa_{(0)}}{3}} y_{(1)} y_{(3)}
\right) \;.
\eea

\subsection{Correlation functions}
\label{subsec:correlation}

\paragraph{One-point functions}

From the renormalized action (\ref{Sren}) we may now extract the one-point correlation 
functions for the various dual operators by functional derivation.  For the operators dual to the
dilaton and the two-dimensional metric, we thus obtain
\begin{equation}
\begin{aligned}
  \langle {\cal{O}}_{\kappa} (t) \rangle 
&=  \frac{1}{{|f_{(0)}(t)|}^{1/2}} \, \frac{\delta S_{\text{ren}}}{\delta \kappa_{(0)}(t)} 
=  \,e^{\kappa_{(0)}} \,\left( - \frac{7}{9} \kappa_{(7)} -
 \frac{22}{45} x_{(2)} x_{(5)}  - \frac{1}{9} e^{-\frac{2 \kappa_{(0)}}{3}} y_{(1)} y_{(3)}
\right) \,,\\
  \langle {\cal{O}}_{f} (t) \rangle 
&=  \frac{2}{{|f_{(0)}(t)|}^{1/2}} \, \frac{\delta S_{\text{ren}}}{\delta f_{(0)}^{-1}(t)}
=e^{\kappa_{(0)}} \,  \left(  \frac{7}{9} \kappa_{(7)} +
 \frac{22}{45} x_{(2)} x_{(5)}  + \frac{1}{3} e^{-\frac{2 \kappa_{(0)}}{3}} y_{(1)} y_{(3)}
\right) \,.\\
\end{aligned}
\end{equation}
Similarly, in the matter sector, we derive the following
one-point correlation functions for the operators dual to the 
scalars in the {\bf 44} and the {\bf 84} representation
\begin{align}
\langle {\cal{O}}_{44} (t) \rangle 
&=  \frac{1}{{|f_{(0)}(t)|}^{1/2}} \, \frac{\delta S_{\text{ren}}}{\delta x_{(2)}(t)} 
\propto e^{\kappa_{(0)}} \, x_{(5)}(t)\,,\\
\langle {\cal{O}}_{84} (t) \rangle 
&=  \frac{1}{{|f_{(0)}(t)|}^{1/2}} \, \frac{\delta S_{\text{ren}}}{\delta y_{(1)}(t)} 
\propto   e^{\kappa_{(0)}/3} \, y_{(3)}\,.
\end{align}

\paragraph{Two-point function}

The two-point correlation functions are obtained by further functional derivative of the one-point functions.
To this end, we first need to determine the dependence of the `response' functions 
$\{f_{(7)}, \kappa_{(7)}, x_{(5)}, y_{(3)}\}$ on the `source functions' $\{f_{(0)}, \kappa_{(0)}, x_{(2)}, y_{(1)}\}$.
This dependence is fixed by the requirement that the solution of the field equations remains regular in the bulk. 
In absence of an exact solution of the non-linear equations of motion, the two-point correlation functions 
can be computed from exact solutions of the linearized equations of motion.

In the dilaton-gravity sector, linearizing the field equations around the background 
\begin{align}
 f(t,r) &= 1+ \eta(t,r) \,,\nonumber\\
\kappa(t,r) &= 0 + \kappa(t,r)\,,
\end{align}
leads to the set of equations
\begin{align}
0&=\frac{1}{2} \eta^{''} + \kappa^{''} \,,\qquad
0=\dot{\kappa}^{'} \,,\nonumber\\
0&=2 \alpha \gamma \, \eta^{'} + 2 r \, \eta^{''} + \ddot{\kappa} - 2 \kappa^{'} \,,\nonumber\\
0&=4r \kappa^{''} + \left( 2+ 8  \alpha \gamma\right) \kappa^{'} + \ddot{\kappa} + 2 \alpha \gamma \, \eta^{'}\,,
\end{align}
whose general solution is provided by 
\begin{align}
\label{linG}
\eta(t,r) &= \eta_{(0)}(t) +\frac59\,\ddot{\kappa}_{(0)}(t) \,  r  - 2 A \, r^{7/5} \,,\nonumber\\
\kappa(t,r)&= \kappa_{(0)}(t) + A \, r^{7/5}  \,,
\end{align}
with real constant $A$. 
Regularity in the bulk requires that $A=0$ which translates into $f_{(7)}=0=\kappa_{(7)}$\,.
As a result, all related two-point correlation functions vanish.
\begin{equation}
 \langle {\cal O}_{\kappa}(t_1)  {\cal O}_{\kappa}(t_2)  \rangle =   0 =
 \langle {\cal O}_{f}(t_1)  {\cal O}_{f}(t_2)  \rangle 
  \;.
\end{equation}
As alluded to above, this is a consequence of the fact that in two dimensions
the dilaton-gravity sector does not carry propagating degrees of freedom.

The interesting structure of correlation functions is situated in the matter sector. 
Linearizing the scalar field equations (\ref{eomy}) around the background (\ref{dw11})
yields  a linear differential equation that can be simplified by taking the Fourier transform with respect to time:
\begin{equation}
\label{scalar}
r^{2} \, \tilde{y}_n''(q,r) + \Big(\frac{21}{20} \, a_n - \frac{2}{5} \Big) r \, \tilde{y}_n'(q,r) - \frac{1}{4} ( q^{2} r - m_n^2 ) \, \tilde{y}_n(q,r)=0
\;.
\end{equation} 
For the scalars from the {\bf 44} and the {\bf 84} with the parameters given by (\ref{amy}), the asymptotic analysis of this equation
yields an expansion
\bea
\tilde{y}_{(44)}(r,q) &=& r^{2/5} \left(\tilde{x}_{(2)}(q) + r^{3/5} \, \tilde{x}_{(5)}(q) +\dots\right)
\;,\nonumber\\
\tilde{y}_{(84)}(r,q) &=& r^{1/5} \left(\tilde{y}_{(1)}(q) + r^{2/5} \, \tilde{y}_{(3)}(q) +\dots\right)\;,
\label{axy}
\eea
in accordance with (\ref{ax}).

Let us first
consider the scalar fields in the {\bf 44}. The corresponding equation (\ref{scalar}) can 
 be brought in a more canonical form by making the following change of variables and redefinitions
\begin{equation}
\label{can}
  \tilde{r} = q\, \sqrt{r}\,, \qquad
 \tilde{y}_{(44)}(q,\tilde{r}) = \tilde{r}^{\lambda} \, s(q,\tilde{r}) \,, \qquad
\lambda = \frac{7}{5}\,,
\end{equation}
upon which the equation becomes
\begin{equation}
 \label{scan}
\tilde{r}^{2} s^{''} + \tilde{r} \, s^{'} -\big(\tilde{r}^2 + \lambda^2 -m^2\big) \, s =0\,.
\end{equation}
This corresponds to the modified Bessel's equation with parameter $\sqrt{\lambda^2-m^2}=\frac35$. 
It admits two linearly independent solutions which may be described by
 modified Bessel function of the first kind $I$ and the second kind $K$. 
 The solution regular in the bulk is given by
\bea
\tilde{y}_{(44)}(q,r) &=&   \tilde{r}^{7/5}  \, \mathrm{Bessel}_K (3/5,\tilde{r})\,,
\label{Bess}
\eea
and we can infer its asymptotic development near $r=0$ as
\begin{align}
 \tilde{y}_{(44)}(q,r) &=q^{4/5}   \, \left(  \frac{\Gamma (\frac{3}{5})}{2^{2/5}} \,  r^{2/5} +  \frac{ \Gamma(-\frac{3}{5})}{ 2^{8/5}} \, q^{6/5} \, r
+ \frac{5 \Gamma(\frac{3}{5})}{ 2^{17/5}} \, q^2 \, r^{7/5} + \underset{r \to 0}{o}(r^{7/5}) \right)\;.
\end{align}
Comparing to the general expansion (\ref{axy}) we find that
\begin{equation}\label{251}
{\tilde{x}}_{5}(q) \propto q^{6/5}\, {\tilde{x}}_{2}(q)  \,.
\end{equation}
Before proceeding with the computation of the two-point function, we should recall the 
possible ambiguity in the assignment of conformal dimensions for the scalar fields discussed in 
appendix~\ref{app:amb}.
The scalar fields in the {\bf 44} precisely live in the mass range that allows for two different field
theory interpretations.
On the level of the present discussion, the two different choices simply correspond to an exchange of
the role of `source' and `response' function ${\tilde{x}}_{2}(q)$ and ${\tilde{x}}_{5}(q)$~\cite{Klebanov:1999tb}.

Accordingly, the two-point function in momentum space is given by
\begin{equation}
\label{44ex}
\langle {\cal{O}}_{44} (0) {\cal{O}}_{44} (q) \rangle \propto q^{\pm6/5} \,,
\end{equation}
and after Fourier transformation
\begin{equation}
\label{corr1}
\langle {\cal{O}}_{y} (t_{1}) {\cal{O}}_{y} (t_{2}) 
\rangle \propto \text{TF}^{-1}(q^{\pm6/5})(t_{1}-t_{2}) \propto \frac{1}{|t_{1}-t_{2}|^{1\pm (6/5)}} \,.
\end{equation}

For the scalars in the {\bf 84}, equation (\ref{scalar}) turns into a Bessel equation (\ref{scan}) with $\lambda=\frac45$, 
such that its regular solution is given by
\begin{equation}
 \tilde{y}_{(84)}(q,r) = \tilde{r}^{4/5} \, \mathrm{Bessel}_K (2/5,\tilde{r})\,,
\end{equation}
with near $r=0$ series expansion
\begin{equation}
 \tilde{y}_{(84)}(q,r) = q^{2/5}  \, \left( \frac{\Gamma (\frac{2}{5})}{2^{3/5}} \, r^{1/5} + 
 \frac{\Gamma (-\frac{2}{5})}{ 2^{7/5}}\; q^{4/5}  \, r^{3/5}
+ \frac{5 \Gamma (\frac{2}{5})}{12\ 2^{3/5}} \; q^2 \, r^{6/5} + \underset{r \to 0}{o}(r^{6/5}) \right)\,.
\end{equation}
Thus, the first two coefficients in the expansion (\ref{axy}) are related by
\begin{equation}
{\tilde{y}}_{3}(q) \propto q^{4/5} \,{\tilde{y}}_{1}(q)  \,.
\end{equation}
Again depending on the choice of assigment $\Delta_{\pm}$, 
the two-point function is thus given by
\begin{equation}
\label{corr2}
\langle {\cal{O}}_{84} (t_{1}) {\cal{O}}_{84} (t_{2})
 \rangle \propto \text{TF}^{-1}(q^{\pm4/5})(t_{1}-t_{2}) \propto \frac{1}{|t_{1}-t_{2}|^{1\pm(4/5)}}\,.
\end{equation}

\subsection{Comparison to the matrix model}\label{sec:25}

The dual field theory is the super matrix quantum mechanics,
obtained by dimensional reduction of ten-dimensional SYM theory
to one dimension, where it is of the form~\cite{deWit:1988ig}
\bea
{\cal L}_{\rm MQM} &=& {\rm tr}\left\{
(D_t \,{\bf X}^k)^2 + \psi^I D_t \psi^I
- \frac12[\,{\bf X}^k,\,{\bf X}^l]^2 - \Gamma^k_{IJ}\, \psi^I [\,{\bf X}^k, \psi^J] 
\right\}
\;,
\label{sym}
\eea
with $SU(N)$ valued matrices $\,{\bf X}^k$, $\psi^I$ in the corresponding vector and spinor representations 
of $SO(9)$. This model
itself has been proposed as a non-perturbative definition of
M-theory~\cite{Banks:1996vh}. 
The gauge invariant operators dual to the supergravity scalars in the {\bf 44} and the {\bf 84}, respectively,
can be identified via their $SO(9)$ representations
\bea
\label{op2}
{\cal O}_{44} &\propto& T_{ij}^{++}~=~ \frac{1}{N} \Big( \text{tr} \big( \,{\bf X}^i \,{\bf X}^j \big) - \frac{1}{9}\,\delta^{ij}\,
 \sum_{k=1}^9  \text{tr} \big( \,{\bf X}^k \,{\bf X}^k \big) \Big)\,,
\nonumber\\
{\cal O}_{84} &\propto&J_{ijk} ~\propto~
\frac{1}{N}\, \text{tr} \big( [\,{\bf X}^i,\, \,{\bf X}^j] \,{\bf X}^k \big)
\;,
\eea
The behaviour of these operators in the matrix quantum mechanics has been studied in 
\cite{Hanada:2011fq} by Monte Carlo methods. Their result shows precise agreement with
(\ref{corr1}) and (\ref{corr2}) if we select $\Delta_-$ for the {\bf 44} scalars and $\Delta_+$
for the {\bf 84} scalars, respectively. 
Only this assignment will correspond to a supersymmetric field theory dual.
This result also agrees with the linearized Kaluza-Klein analysis 
of~\cite{Sekino:1999av} (where the issue of the $\Delta_\pm$ ambiguity was not discussed).
In the next section we will use the full non-nonlinear effective theory in order to compute
correlation functions for deformations of the model~(\ref{sym}).

\section{Deformed BFSS model holography}
\label{sec:deformation}

In the following section 
we will construct a half-supersymmetric `deformed' domain-wall solution of two-dimensional $SO(9)$ supergravity which, as it turns out, uplifts to an eleven-dimensional pp-wave with $SO(3)\times SO(6)$ symmetry. We will see  however that the resulting eleven-dimensional pp-wave does not belong to the class of  bubbling  M-theory $SO(3)\times SO(6)$ geometries  of \cite{Lin:2004nb}. In particular, contrary to \cite{Lin:2004nb}, our eleven-dimensional pp-wave background has vanishing four-form flux and is consistent with the analysis of \cite{OColgain:2012wv}. From its asymptotic behaviour we conclude that it describes
a vev deformation of the BFSS matrix model.
In sections \ref{sec:32}, \ref{sec:33} we then use holographic renormalization as developed in the last section
to compute around this solution two-point correlation functions of operators dual 
to the ${\bf 44}$ scalar fields which decompose into
\bea
{\bf 44} &\longrightarrow (1,20)\oplus(5,1)\oplus(3,6)\;,
\label{44bmn}
\eea
under $SO(3)\times SO(6)$.

\subsection{$SO(3) \times SO(6)$ domain wall}

In this section, we determine the half-maximal BPS solutions of the maximal two-dimensional 
supergravity~(\ref{Ltrunc}) that preserve an $SO(3)\times SO(6)\subset SO(9)$ subgroup of the gauge symmetry.
A simple ansatz for such a vacuum solution is provided by exciting the 
scalars 
\begin{equation}
\label{s3s6trunc}
 X_{1,2,3} = e^{-x}\;,\quad X_4 = e^{2x}\;,
\end{equation}
and setting the axion fields $\phi_a$ to zero.
 The $SO(3)\times SO(6)$ symmetry can be easily seen from the embedding of the $U(1)^4$ truncation (\ref{Ltrunc}) into the full $SO(9)$ theory \cite{Ortiz:2012ib}, where the $SL(9)/SO(9)$ coset space is parametrized by an $SL(9)$ valued scalar matrix ${\cal V}$. 
In the  $U(1)^4$ truncation this matrix is diagonal
\begin{equation}
 {\cal{V}} = \text{diag}\, \big( X_{1}^{-1/2},\,X_{1}^{-1/2},\, \dots ,\, X_{4}^{-1/2},\,X_{4}^{-1/2},\, X_1 X_2 X_3 X_4  \big)\,.
\end{equation}
With the ansatz \eqref{s3s6trunc}, it takes the form
\begin{equation}
 {\cal{V}}= \begin{pmatrix}
e^{x/2} \mathbb{I}_{6\times 6} & 0 \\
0 & e^{-x} \mathbb{I}_{3\times 3} \\
\end{pmatrix}
\,,
\label{V36}
\end{equation}
which preserves an $SO(3)\times SO(6)$ subgroup of the $SO(9)$ gauge symmetry. 
The two-dimensional bosonic effective Lagrangian (\ref{Ltrunc}) becomes
\begin{equation}
\label{leff2}
{\cal{L}} = -\frac{1}{4} e  \rho  R 
+ \frac{9}{8} e \rho \, ( \partial_{\mu}x) ( \partial^{\mu} x)
+ \frac{3}{8} e \rho^{5/9} \, e^{-2x} \, (8 +12 e^{3x} + e^{6x})\,.
\end{equation}
In the following we will construct BPS solutions in this truncation of the theory. We stress that 
the $U(1)^4$ truncation (\ref{Ltrunc}) is presumably not the bosonic sector of a supersymmetric theory
but can be embedded into the maximally supersymmetric $SO(9)$ theory of \cite{Ortiz:2012ib}, which allows to 
discuss BPS solutions of the latter. The full theory has 16 gravitinos, 16 dilatinos and
128 fermions. Vanishing of their supersymmetry transformations in the truncation \eqref{s3s6trunc}
implies the Killing spinor equations
\begin{align}
 0 &\overset{!}{=}  \partial_{\mu} \epsilon^I + \frac{1}{4} {\omega_{\mu}}^{\alpha \beta} \gamma_{\alpha \beta} \epsilon^I
+ \frac{7}{12} \,i \rho^{-2/9} (e^{2x}+2e^{-x})\, \gamma_{\mu} \epsilon^I \,,\nonumber\\ 
 0 &\overset{!}{=}  - \frac{i}{2} (\rho^{-1} \partial_{\mu} \rho)\,  \gamma^{\mu} \epsilon^I + \frac{3}{4} \rho^{-2/9} (e^{2x}+2e^{-x})\, \epsilon^I\,,\nonumber\\
0 &\overset{!}{=}(\partial_{\mu} x)\, \gamma^{\mu} \epsilon^I - \frac{2i}{3} \rho^{-2/9} (e^{2x}-e^{-x})\, \epsilon^I\,,
\label{KSE}
\end{align}
for the Killing spinor $\epsilon^I$, $I=1, \dots, 16$\,. Here, $\omega_\mu{}^{\alpha\beta}$ is the spin connection and
$\gamma_\alpha$ denote the $SO(1,2)$ gamma matrices.
Apart from the $SO(9)$ invariant solution (\ref{dw}) for which $x=0$, 
these equations admit a unique non-trivial solution. Part of the diffeomorphism invariance 
can be fixed upon identifying the scalar $x$ with the radial coordinate, after which the solution takes the form
\bea
\label{backbmn}
\rho(x) &=& e^{\frac{9}{2} x} (e^{3x}-1)^{-9/4}\;,\qquad
ds_{2}^2 ~=~ \widetilde{f}(x)^2 dt^2 - \widetilde{g}(x)^2 dx^2\,,
\eea
with the functions
\bea
\widetilde{f}(x) &\equiv&  e^{\frac{7}{2} x}(e^{3x}-1)^{-7/4}\,,\qquad
\widetilde{g}(x) ~\equiv~\frac{3}{2}  e^{2  x} (e^{3x}-1)^{-3/2}\,,
\eea
up to coordinate redefinitions.
The associated Killing spinors are given by
\bea
\epsilon^{I}(x) &= a(x) \, \epsilon_{0}^{I}\;,\qquad\mbox{with}\quad
\gamma^1 \epsilon_{0}^{I} = -i \epsilon_{0}^{I}\;,
\label{KS}
\eea
and a function $a(x)$ that is obtained from integrating the first equation of (\ref{KSE}).
This confirms that the background preserves sixteen supercharges, i.e.\ has the same number of
supersymmetries as the $SO(9)$ domain wall (\ref{dw}).
Since $x$ is non-vanishing in the bulk, this deformation breaks $SO(9)$ 
down to $SO(3)\times SO(6)$\,.
The Ricci scalar of the two-dimensional metric (\ref{backbmn}) takes the following form
\bea
 R &=& -\frac{5}{6} \,  e^{-2x} \big( e^{6x} -12 e^{3x} - 4 \big)\,, \nonumber\\[1ex]
 &&{}\mbox{such that}\quad
 R ~=~ \frac{25}{2} \,  + \underset{x \to 0}{{\cal O}}(x^2)\,, \quad 
R ~=~ -\frac{5}{6} \,  e^{4x} + 10\,  e^x + \underset{x \to +\infty}{o(1)} \;.
\eea
It is well defined on the interval $x\in [0\,,\,+\infty[$\, in contrast to the metric  and the dilaton which are singular at $x=0$.

\paragraph{Higher-dimensional interpretation.}

Although the geometry of the solution (\ref{backbmn}) 
may be obscure in this parametrization, its interpretation becomes clearer in eleven dimensions. 
Its uplift to ten dimensions can be performed using the embedding of $SO(9)$ supergravity in type IIA supergravity \cite{Anabalon:2013zka}. 
The resulting solution of the type IIA bosonic equations of motion takes the form
\begin{align}
ds_{10}^2 &= \rho^{-7/36}
\Delta^{7/8} \,ds_2^2 
-  \rho^{1/4}\,\Delta^{-1/8}\,\Big( 
\frac{\Delta}{e^x (1-\mu^2)} \, d\mu^2 + e^{-2x}(1-\mu^2) \, d\Omega_{2}^{2}
+e^x\mu^2 \, d\Omega_{5}^{2} \Big)\;, \nonumber\\
\Phi&= \frac{1}{3} \log \big( \rho^{-7/4} \Delta^{-9/8}\big) \,,\nonumber\\
F &= 2 \rho^{5/9}  \, \big( f_1(x) + \mu^2 \,f_2(x) \big) \, \varepsilon_2
 - \frac{3}{2 } \rho\,( *_2 dx) \wedge d(\mu^2)\equiv dA_1\,,
\end{align}
for metric, dilaton and two-form flux,
where
\begin{align}
0 &\leq \mu^2 \leq1\,, \quad \Delta \equiv e^{2x}+\mu^2 (e^{-x}-e^{2x})\,,\nonumber\\
f_1 (x)&\equiv -\frac{1}{2} e^{2x} (e^{2x}+6e^{-x}) \,,\quad f_2 (x) \equiv 
-\frac{1}{2} (e^{-x}-e^{2x}) (4e^{-x}+e^{2x})\,.
\end{align}
This solution allows straightforward uplift to a purely geometric solution of the $D=11$ Einstein equations
according to
\begin{align}
ds_{11}^{2}&=-2\,dt  dz  - \frac{\left(e^{3 x}-1\right)^{7/2}}{\left(1-\mu ^2\right) e^{9 x}+\mu ^2 e^{6 x}} \, dz^2 
-\frac{1-\mu ^2}{ e^{3 x}-1}\, d\Omega_2^2 -\frac{\mu ^2 e^{3 x}}{ e^{3 x}-1} \, d\Omega_5^2\nonumber\\
&\quad-\frac{9 \, \text{csch}^2\left(\frac{3 x}{2}\right) \left(1-2 \mu ^2+\coth \left(\frac{3
   x}{2}\right)\right)}{32} \, dx^2 - \frac{\left(1-\mu ^2\right) e^{3 x}+\mu ^2}{\left(1-\mu ^2\right)
   \left(e^{3 x}-1\right)}  \, d\mu^2\;.
\end{align}
Eventually, this expression can be considerably simplified by the following coordinate transformations
\bea
r_{3}^{2} = \frac{1-\mu ^2}{ e^{3 x}-1}\,,\qquad  r_{6}^{2} = \frac{\mu ^2 e^{3 x}}{ e^{3 x}-1}\,,\qquad
x^{\pm} = t\pm(t+z)\,,
\eea
upon which the metric becomes
\begin{equation}
ds_{11}^{2} =  dx^{+} \, dx^{-} - H(r_3,r_6) (dx^{-})^2 
-  \Big(dr_{3}^{2} + r_{3}^{2}\, d\Omega_2^2 + dr_{6}^{2} + r_{6}^{2}\, d\Omega_5^2 \Big) \,,
\label{dw11}
\end{equation}
where the function $H(r_3,r_6)$ is given by
\bea\label{316}
H(r_3,r_6)&\equiv&
\frac{(1+\gamma-r_3^2-r_6^2)^{\frac52}(1+\gamma+r_3^2-r_6^2)^{-2}}{\sqrt2\,\gamma\,r_3}
\;,\nonumber\\
&&\nonumber\\
\gamma&\equiv& \sqrt{(1 + r_3^2 + r_6^2 + 2 r_6 )(1 + r_3^2 + r_6^2 - 2 r_6 )}
\;.
\eea
Remarkably (but necessarily for consistency) $H(r_3,r_6)$ 
satisfies the Laplace equation $\Delta H=0$ on the Euclidean space $\mathbb{E}^9$. Consequently the metric (\ref{dw11}) represents
 a pp-wave solution of eleven-dimensional supergravity \cite{Gauntlett:2002cs}. Just as the domain-wall solution \eqref{2.6}, 
 it is a purely gravitational solution in eleven dimensions.

From the ten-dimensional point of view the solution can in fact be interpreted as the near-horizon limit 
of a distribution of D0 branes with $SO(3)\times SO(6)$ symmetry, similarly to the multi-centered solutions of \cite{Freedman:1999gk} for D3 branes\footnote{We are grateful to the referee of JHEP for bringing up this point.}. 
To make the form of the distribution explicit, note that $H(r_3,r_6)$ in (\ref{316}) takes the form,
\begin{equation}
H(r_3,r_6)\sim\frac{1}{r_3}(1-r_6^2)^{-\frac12}+\mathcal{O}(r_3)
~.
\end{equation}
This suggests that the D0 branes are localized at $r_3=0$ in three of the nine transverse dimensions, while they follow a distribution given by
\begin{equation}
\sigma(r_6)=\left\{\begin{array}{cr}
(1-r_6^2)^{-\frac12}&~,~~ r_6<1\\
0 &~,~~ r_6\geq 1
\end{array}\right.
~,
\end{equation}
in the remaining six transverse dimensions. Indeed it can be checked by a direct calculation  that 
\begin{equation}
H(r_3,r_6)=\frac{15\sqrt{2}}{2\pi^3}\int\mathrm{d}^9y~\!\delta(\vec{y}_3)~\!\sigma(|\vec{y}_6|)~\!\frac{1}{|\vec{x}_9-\vec{y}_9|^7}
~~\!,
\end{equation}
where the position vector $\vec{x}_9$ in the transverse directions splits as $\vec{x}_9=\vec{x}_3+\vec{x}_6$ with $r_3:=|\vec{x}_3|$, $r_6:=|\vec{x}_6|$.

\paragraph{Operator vs. vev deformation.}

Let us consider the 1/2-BPS solution \eqref{backbmn}. After going to the Euclidean signature and making the following Weyl rescaling 
\begin{equation}
g_{\mu \nu} \;\rightarrow\; \rho^{4/9} \, g_{\mu \nu} \;,
\label{W49}
\end{equation}
and coordinate change $(x= r^{2/5})$, one recovers the metric of an asymptotically AdS spacetime coupled to a dilaton:
\bea
\label{backW2}
d{\widehat{s}_{2}}^2 &=& \widehat{f}(r)^2 dt^2 + \widehat{g}(r)^2 dr^2 
\;,\nonumber\\
&&}{ \widehat{g}(r) \equiv \frac{3}{5} x^{-3/2} \,  e^{x} (e^{3 x} - 1)^{-1}\,,\quad
\widehat{f}(r) \equiv 3^{5/4} \, e^{\frac{5}{2} x} (e^{3 x}-1)^{-5/4}\,.
\eea
Indeed, up to some global numerical constants, in the limit $(r \to 0)$
one recovers the dilaton coupled AdS background \eqref{backG} 
 \bea
 d{\widehat{s}_{2}}^2 &\underset{r \to 0}{\sim}& \frac{dt^2}{r} + \frac{dr^2}{4  r^2} \;,
 \qquad \rho(t,r) ~\underset{r \to 0}{\sim}~ r^{-9/10}\,.
 \eea
 In this frame where the metric is asymptotically AdS, the near boundary behavior of the scalar field $x(r)$ 
 allows to identify whether the gauge theory dual to the 1/2-BPS solution~\eqref{backbmn} corresponds
 to an operator deformation or a vev deformation of the undeformed BFSS matrix 
 model~\cite{Klebanov:1999tb,Skenderis:2002wp}. 
Recall that the correct near-boundary asymptotic form for a scalar $\phi$ propagating in the 
AdS$_{d+1}$ bulk which is dual to a dimension-$\Delta$ operator in the boundary CFT is given by:
\begin{equation}
\phi=r^{d-\Delta}\varphi_s+\dots+r^{\Delta}\varphi_v+\dots~\;.
\label{asympP}
\end{equation}
Via the AdS/CFT dictionary  $\varphi_s$ is the source for the CFT operator dual to $\phi$, while $\varphi_v$ is its vev
(unless the conformal dimension $\Delta$ is in the critical interval which allows for an interchange of the interpretation,
as reviewed in appendix~\ref{app:amb}).

If instead of an AdS$_{d+1}$ bulk we have an asymptotically  AdS$_{d+1}$ geometry which is supported by a nontrivial profile for the bulk field $\phi$ above, we can have two possible scenarios corresponding to two different deformations of the gauge theory:

\begin{itemize}

\item Operator deformation: this corresponds to an asymptotic behavior $\phi\sim r^{d-\Delta}\varphi_s$ near the boundary.

\item Vev deformation: this corresponds to an asymptotic behavior $\phi\sim r^{\Delta}\varphi_s$ near the boundary.

\end{itemize}

With the general expansion of the active scalar field from (\ref{ax})
\bea
y_{44}(r,t)&= r^{2/5} \, x_{(2)}(t) +  r \, x_{(5)}(t)  + \dots
\;,
\label{asyy}
\eea
we find that around $r=0$, the background \eqref{backW2}
\begin{equation}
 x(r) = r^{2/5}\;,
 \label{asyx}
\end{equation}
corresponds to the first term in \eqref{asympP}. 
However, as we have discussed after (\ref{op2}) above, the BFSS matrix model corresponds to the opposite
choice $\Delta_-$ of conformal dimension for the scalar fields in the ${\bf 44}$. I.e.\ the role of source
and response in (\ref{asympP}) are exchanged and an asymptotic behavior (\ref{asyx}) of the active scalar field
implies the holographic interpretation as a vev deformation.
We conclude that the holographic dual to the background \eqref{backbmn} corresponds to a vev deformation of the BFSS model \cite{Skenderis:2002wp}. 
A domain wall with opposite boundary behaviour on the other hand would describe 
an operator deformation of the BFSS model such as the BMN matrix model~\cite{Berenstein:2002jq}.
The corresponding gravitational background presumably requires also non-vanishing axion fields.
In the following, we will compute correlation functions in the deformed matrix model from the gravity side and 
interpret them in the light of the gauge/gravity correspondence.

\subsection{On-shell action and Renormalization}
\label{sec:32}

The procedure to compute holographic correlation functions around the background \eqref{backbmn}
is the same which we have followed in section~\ref{sec:BFSS} for the correlation functions of the BFSS model.
As the first step, we will compute the effective action
 that describes scalar fluctuations around the background \eqref{backbmn}.

\subsubsection{Effective action}
 We will study fluctuations of the full $SO(9)$ supergravity around the background \eqref{backbmn}.
To this end we consider the $SL(9)$ valued matrix ${\cal V}$. Its fluctuations are
most conveniently expressed by a parametrization 
\begin{equation}
 {\cal{V}} \equiv {\cal{V}}_{\text{background}} \, \Big( \mathbb{I}_{9\times9} + X + \frac{1}{2} X^2 + \dots \Big)
 \;,
\end{equation}
where ${\cal{V}}_{\text{background}}$ corresponds to the matrix (\ref{V36})
evaluated on the background solution, 
and $X \in {\mathfrak{sl}}(9)$ carries the scalar fluctuations. Since the background breaks 
$SO(9)$ down to $SO(3)\times SO(6)$, the fluctuations organize into 
irreducible representations of $SO(3) \times SO(6)$:
\begin{align}
{\bf 44}~\longrightarrow~  (1,1) \oplus (5,1) \oplus (1,20) \oplus (3,6) \,.
\label{scalarfluc}
\end{align}
The perturbations $x_{(5,1)}$ and $x_{(1,20)}$ are already captured by the $U(1)^4$ truncation 
(\ref{s3s6trunc}) and obtained by setting
\begin{equation}
 X_{1,2}=e^{-x+x_{(1,20)}}\,,\quad X_{3}=e^{-x-2x_{(1,20)}} \,,\quad X_4 = e^{2x - 2 x_{(5,1)}}\,.
\end{equation}
In contrast, the fluctuations in the $(3,6)$ do not sit within the $U(1)^4$ truncation so that 
their description requires the full $SO(9)$ theory.
We will not consider in the following the perturbation in the singlet $(1,1)$, since its interaction with the metric
fluctuations leads to rather non-trivial non-diagonal couplings in the action. 
The resulting Euclidean action quadratic in the scalar fluctuations (\ref{scalarfluc}) is given by
\begin{align}
S &=- \int dx^2\, e \, \Bigg( -\frac{1}{4}  \rho  R 
+ \frac{9}{8} e \rho \, ( \partial_{\mu}x) ( \partial^{\mu} x)
- \frac{3}{8} e \, \rho^{5/9} \, e^{-2x} (8 +12 e^{3x} + e^{6x})\nonumber\\
&\quad\qquad\qquad+ \frac{1}{2}\,e \rho (\partial x_{(5,1)})^2 + e \, \rho^{5/9} \, e^x (e^{3x}-6) \, x_{(5,1)}^2 \nonumber\\
&\quad\qquad\qquad + \frac{1}{2}\,e \rho  (\partial x_{(1,20)})^2 - e \, \rho^{5/9} \,(2 e^{-2x}+ 3 e^x) \, x_{(1,20)}^2 \nonumber\\
&\quad\qquad\qquad 
+\frac{1}{2}\,e \rho  (\partial x_{(3,6)})^2 - e \,  \rho^{5/9} \, \frac{e^{-2x}}{2}(3 + 5 e^x + 2 e^{3x}) \, x_{(3,6)}^2 \Bigg)\,.
\end{align}
As we have seen above, the renormalization process is more easily done after the Weyl rescaling (\ref{W49})
upon which the dilaton enters the action as a global factor. 
In this frame, the effective action becomes
\bea
\label{SE}
S &=& \frac{1}{4} \int d^2 x \, e  \rho \, \Big(  R + \frac{4}{9} \big( \rho^{-1} \partial \rho \big)^2
- \frac{9}{2}  \, ( \partial_{\mu}x) ( \partial^{\mu} x)
+ \frac{3}{2}   \, e^{-2x} (8 +12 e^{3x} + e^{6x})\nonumber\\
&&{}\qquad- 2\, (\partial x_{(5,1)})^2 
 -2\,  (\partial x_{(1,20)})^2 -2\,  (\partial x_{(3,6)})^2   - 4  \, e^x (e^{3x}-6) \, x_{(5,1)}^2
 \nonumber\\
 &&{}\qquad
  + 4    \,(2 e^{-2x}+ 3 e^x) \, x_{(1,20)}^2   
  + 2  \, e^{-2x}(3 + 5 e^x + 2 e^{3x}) \, x_{(3,6)}^2 \Big)\,.
 \eea
The associated equations of motion are given by
\begin{align}
0&=\rho^{-1} \nabla \partial \rho - \frac{3}{2}\, e^{-2 x}(8 +12 e^{3x} + e^{6x})
-  \sum_{i\in I} F_{i}(x) \, x_{i}^2  \,,\nonumber\\[1ex]
0&= \rho^{-1} \big(\nabla_{\mu} \partial_{\nu} \rho - \frac{1}{2} g_{\mu \nu} \nabla \partial \rho \big)
-\frac{4}{9}\, \rho^{-2} (\partial_{\mu}\rho \partial_{\nu}\rho -\frac{1}{2} g_{\mu \nu} (\partial \rho)^2)
+\frac{9}{2}\, \big( \partial_{\mu} x \partial_{\nu} x - \frac{1}{2}\, g_{\mu \nu} (\partial x)^2 \big)  \nonumber\\
&\qquad+2 \sum_{i\in I} \big( \partial_{\mu} x_i \partial_{\nu} x_i - \frac{1}{2}\, g_{\mu \nu} (\partial x_i)^2 \big) \,,\nonumber\\[1ex]
0&= R + \frac{4}{9}\, \rho^{-2} (\partial \rho)^2  - \frac{8}{9} \rho^{-1} \nabla \partial \rho
- \frac{9}{2} (\partial x)^2 +\frac{3}{2}\, e^{-2 x}(8 +12 e^{3x} + e^{6x}) \nonumber\\
&\qquad-2  \, \sum_{i\in I} \big((\partial x_i)^2- \frac{F_{i}(x)}{2} \, x_{i}^2 \big) \,,
\end{align}
and
\begin{align}
0&= \rho^{-1} \nabla \big(\rho \, \partial x \big) - \frac{2}{3}\,  e^{-2 x}(4 -3 e^{3x} - e^{6x})  
+ \frac{1}{9}\, \sum_{i\in I} F_{i}'(x)\, x_{i}^2   \,,\nonumber\\
0 &=\rho^{-1}\nabla(\rho \, \partial x_i)+ \frac{1}{2}\, F_{i}(x)\, x_{i}\,,
\label{eqmFS}
\end{align}
with $I \equiv \{(5,1)\,,\,(1,20)\,,\,(3,6) \}$, and the scalar functions
\bea
 F_{(5,1)} &=& -4 \, e^x (e^{3x}-6) \,,\qquad F_{(1,20)} ~=~ 4   \,(2 e^{-2x}+ 3 e^x) \,, \nonumber\\ 
F_{(3,6)} &=&2   \, e^{-2x}(3 + 5 e^x + 2 e^{3x}) \,,
\eea
which capture the interactions of the scalar fluctuations with the background $x(t,r)$ from \eqref{backW2}.

\subsubsection{On-shell action and renormalization}

Again, the effective action \eqref{SE} is most conveniently evaluated on-shell using the dilaton field equation. 
As in (\ref{AdSEdW0}) this leads to a contribution located at the boundary of the asymptotically 
AdS spacetime background 
\eqref{backW2},
\begin{equation}
\label{osaE}
  S= \frac{1}{2} \int_{r = \epsilon} dt  \, \sqrt{|h|} \left(\frac{4}{9} \,n^{\mu} \partial_{\mu} \rho + \rho\, K\right) \,. 
\end{equation}
In the following we will treat the different irreducible representations of the scalar fluctuations separately
since they do not mix at the quadratic level.
Accordingly, we parametrize the fluctuations of the gravity sector as
 \begin{align}
f(t,r) &= f_{b}(r) \, (1+ f_i(t,r))\,,\nonumber\\
\rho(t,r) &= \rho_{b}(r) \, (1+ \rho_i(t,r))\,,
 \end{align}
where $f_b$ and $\rho_b $ denote the background \eqref{backW2} 
and the fluctuations \linebreak[0] $\{f_i(t,r),\rho_i(t,r)\}$ are functions of the scalar fluctuations $x_i$ and 
vanish at the horizon. No source is turned on in the dilaton-gravity sector.
The equations of motion for the scalar fluctuations $x_i$ are given by the last equation of (\ref{eqmFS}) and
indicate that a power series expansion in $r$ of the solution should begin with $r^{2/5}$ or $r$, cf.~(\ref{asyy}).
Moreover evaluation of the on-shell action \eqref{osaE} 
on the background shows that the dilaton and extrinsic curvature terms diverge as 
 \begin{align}
\sqrt{|h|} \; n^{\mu} \partial_{\mu} \rho &\underset{r \to 0}{\sim} r^{-7/5}\,,\qquad
\sqrt{|h|} \; \rho\, K ~\underset{r \to 0}{\sim}~ r^{-7/5}\,.
 \end{align}
Thus we 
only need to determine the power series expansions up to order $r^{7/5}$, 
with all the other orders vanishing in the renormalization process.
The equations of motion further constrain the expansions to
\begin{align}
f_i(t,r) &= f_{(4)}(t) \, r^{4/5} + f_{(6)}(t) \, r^{6/5} + f_{(7)}(t) \, r^{7/5}  + \dots\,,\nonumber\\
\rho_i(t,r) &= \rho_{(4)}(t) \, r^{4/5} + \rho_{(6)}(t) \, r^{6/5} + \rho_{(7)}(t) \, r^{7/5}  + \dots\,,\nonumber\\
x_i (t,r) &= x_{i(2)}(t) \, r^{2/5} + x_{i(4)}(t) \, r^{4/5} + x_{i(5)}(t) \, r + \dots 
\end{align}
Explicitly, the coefficients are related by 
\begin{align}
&f_{(4)}(t)= a_4 \, x_{i(2)}(t)^2  \,, &f_{(6)}(t)&= a_{6} \, x_{i(2)}(t)^2 \,,\nonumber\\
&\rho_{(4)}(t)= b_4 \,x_{i(2)}(t)^2 \,,  &\rho_{(6)}(t)&= b_{6} \,x_{i(2)}(t)^2 \,,\nonumber\\
&x_{i(4)}(t) = d_{4} \, x_{i(2)}(t) \,,
&\rho_{(7)}(t) &= -\frac{11440}{9} \, x_{i(2)}(t) x_{i(5)}(t) - f_{(7)}(t) \,,
\end{align}
with the numerical coefficients given by
\begin{equation}
\begin{array}{c|c|c|c|c|c|}
 & a_4 & b_4 &  d_4 & a_{6} & b_{6} \\
\hline
i=(5,1)  & -\frac{175}{9} & -35 & -3360 & 847000 & 1524600 \\
\hline
i=(1,20)  & -\frac{175}{9} & -35 & 4200 & -1001000 & -1801800 \\
\hline
i=(3,6)  & -\frac{175}{9} & -35 & -12180 & 3003000 & 5405400  \\
\hline
\end{array}
\end{equation}
for the different scalar fields.
In particular the coefficients $x_{i(2)}(t)$ and $x_{i(5)}(t)$ are left undetermined in the expansion and should
be interpreted as a source and response for the correlation functions. 

We can now evaluate the on-shell action and renormalize the divergences. The divergences
occurring in the on-shell action \eqref{osaE} in the limit $\epsilon \to 0$ are canceled by two counter-terms
 \begin{align}
 S_{\text{ct1}} &= \frac{2}{9} \int_{r = \epsilon} dt  \, \sqrt{|h|} \, 
 (-\frac92\, \rho -\frac12\, \rho^{1/9}-\frac29\, \rho^{-1/3}) \,,\nonumber\\
 S_{\text{ct2}} &= \frac{2}{9} \int_{r = \epsilon} dt  \, \sqrt{|h|}  \, (\kappa_1\, \rho + \kappa_2\, \rho^{5/9}) \, x_{i}(t,\epsilon)^2 \,,
 \end{align}
which correct the dilaton coupling and the scalar potential, respectively,
with the numerical constants given by
\begin{equation}
\kappa_1~=~  \frac{4}{9} \left(9 a_4 +4 b_4\right) \;,\quad
\kappa_2~=~  \frac{2}{27} \left(27\,a_6+a_4\,(9-36\,d_4)+4\,(3\,b_6+b_4-4\,d_4 b_4)\right) \;.
\end{equation}
Consequently, the renormalized action is given by
\begin{align}
 S_{\text{ren}} &= \lim_{\epsilon \to 0} \left( S_{\text{on-shell}} + S_{\text{ct1}} + S_{\text{ct2}} \right)\nonumber\\
 &\propto  \quad \int dt  \; \big( x_{i (2)}(t) \, x_{i (5)}(t) + \frac{1}{2216}\,\rho_{(7)}(t) \big) \,.  
\end{align}
This expression for the renormalized action is in complete analogy with \eqref{Sren} 
so in principle one could have guessed the result. 
Nonetheless, it is interesting to see that the renormalization
process developed in \cite{Bianchi:2001de,Bianchi:2001kw,Skenderis:2002wp} straightforwardly 
works in all cases. In the last step, the coefficients
$x_{i(2)}(t)$ and $ x_{i(5)}(t)$ should be related by imposing regularity of the solution in the bulk
in order to find the two-point functions by derivation of the action.

\subsection{Correlation Functions}
\label{sec:33}

The computation of correlation functions now proceeds completely in parallel with section~\ref{subsec:correlation}.
Let us focus on the scalar two-point functions. They will be generated by the following action
 \begin{align}
  S_{\text{ren}} &\propto  \int dt  \;  x_{i (2)}(t) \, x_{i (5)}(t)\nonumber\\
&\propto \int dq \;  \tilde{x}_{i (2)}(q) \, \tilde{x}_{i (5)}(q) \,,
 \end{align}
where the functions of the momentum $q$ stand for the coefficients of the Fourier transform of $x_i$. 
Regularity in the bulk imposes
a relation between these two coefficients
\begin{equation}
 \tilde{x}_{i (5)}(q) = C_i (q) \, \tilde{x}_{i (2)}(q) \,,
\end{equation}
in analogy with (\ref{251}). The two-point function will be given by
\begin{equation}
\label{cpm} \langle {\cal O}_i(0) {\cal O}_i(q) \rangle \propto C^{\pm1}_i (q) \,,
\end{equation}
where the plus, minus sign in the exponent should be chosen depending on whether 
the source is identified with 
 $\tilde{x}_{i (2)}(q)$, 
 $\tilde{x}_{i (5)}(q)$,  respectively. 
In accordance with the discussion of section \ref{sec:25}, the source in the 
deformed BFSS model should be identified  $\tilde{x}_{i (5)}(q)$; 
this then corresponds to selecting the minus sign in (\ref{cpm}).

In the following subsection the function $C_i$ is determined for each scalar perturbation.
Unlike for the correlation functions in the undeformed matrix model, we can no longer provide
analytical solutions to the scalar fluctuation equations but have to resort to numerical methods 
to determine the functions $C_i$.

\subsubsection{Analytics}

The scheme for calculating the two-point functions is now well defined, cf.\ section~\ref{subsec:correlation}.: 
the first step consists of solving the equations of motion for the scalar perturbations
linearized around the background \eqref{backW2}. After taking the Fourier transform with respect to 
time, we are left with an ordinary second order differential equation in the radial coordinate $r$. 
There exists a unique solution that is regular in the bulk (i.e.\  falls off sufficiently fast as $r$ goes to infinity). 
The power series expansion of this regular solution near the horizon $r=0$ 
allows to compute the ratio 
\bea 
 C_i (q)&\equiv& \frac{\tilde{x}_{i (5)}(q)}{\tilde{x}_{i(2)}(q)} \,,
 \label{ratioC}
\eea
which describes the two-point function of the dual operators.
For computational convenience, we will make the change of variable and field redefinition
\begin{equation}
u =\sqrt{ e^{3 (r^{2/5})}-1} \,,\qquad \tilde{x}_i (u) \rightarrow u^2 \, \tilde{x}_i (u)\,.
\end{equation}
The fluctuation equations then translate into
\begin{align}
0&= \tilde{x}_{(5,1)}''(u) + \frac{2}{u}\big(\frac{2 u^2 -1}{u^2+1}\big)\,\tilde{x}_{(5,1)}'(u)
- \frac{ q^2\,u^3}{(u^2+1)^3}\,\tilde{x}_{(5,1)}(u)\,,\label{flucxxx}\\[1ex]
0&= \tilde{x}_{(1,20)}''(u) + \frac{2}{u}\big(\frac{2 u^2 -1}{u^2+1}\big)\,\tilde{x}_{(1,20)}'(u)
+ \frac{2u^4- q^2\,u^3-2}{(u^2+1)^3}\,\tilde{x}_{(1,20)}(u)\,,\nonumber\\[1ex]
0&= \tilde{x}_{(3,6)}''(u) + \frac{2}{u}\big(\frac{2 u^2 -1}{u^2+1}\big)\,\tilde{x}_{(3,6)}'(u)\nonumber\\
&\quad+ \frac{2 u^6-q^2 u^5-4 u^4-11 u^2-5+5 (u^2+1)^{1/3} u^2+5u^2(u^2+1)^{1/3}}{u^2\,(u^2+1)^3}\,\tilde{x}_{(3,6)}(u)\,,\nonumber
\end{align}
for the different species of scalar fields.
All solutions admit an expansion
\begin{equation}
 \tilde{x}_i (q,u)= \alpha(q) + \beta(q) \, u^3 + \underset{u \to 0}{{o}}(u^3)\,,
\end{equation}
 at $u=0$ (corresponding to $r=0$),
and the ratio (\ref{ratioC}) is given by
\bea
 C_i &\propto& \frac{\beta(q)}{\alpha(q)} \;.
 \label{ratioC1}
\eea

\subsubsection{Numerics}

Unlike for the undeformed matrix model, where the regular solution of the scalar fluctuation equations
could be found in analytical form (\ref{Bess}), the equations (\ref{flucxxx}) can only be solved numerically.
In order to directly extract the ratio (\ref{ratioC1}) of series coefficients in the expansion around $u=0$,
we implement a procedure similar to~\cite{Berg:2001ty,Berg:2002hy}.
To begin, let us introduce another function 
\begin{equation}
 y(q,u) = \tilde{x}(q,u) + \frac{1}{3u} \frac{d\tilde{x}}{du}(q,u)\,,
\end{equation}
whose power expansion around $u=0$ goes as
\begin{equation}
 y(q,u)= \alpha(q) + \beta(q) \, u + \underset{u \to 0}{{o}}(u^3) \,.
\end{equation}
For each perturbation, the corresponding equation of motion for $y$ can be solved numerically
for given initial conditions at $u=0$.
Let $y_1$ and $y_2$ denote the unique solutions with initial conditions 
\begin{equation}
 \{\,y_1(0)=1\,,\; y_{1}'(0)=0\,\}\,, \qquad
\{\, y_2(0)=0\,,\; y_{2}'(0)=1\,\}\,,
\end{equation}
respectively, then the unique solution $y_s$ regular in the bulk (when $u\to +\infty$) 
may be written (up to a global normalization factor) as a linear combination:
\begin{equation}
 y_s = y_1 + \kappa(q) \, y_2 \;= 1 + \kappa(q) \, u + \underset{u \to 0}{{o}}(u^3)
= 1 +\frac{\beta(q)}{\alpha(q)}\, u + \underset{u \to 0}{{o}}(u^3)\,.
\end{equation}
Since $y_1$ and $y_2$ both have the same asymptotic behaviour in the bulk while the combination
$y_s$ vanishes, we may read off the quotient $\beta(q)/\alpha(q)$ from the limit
\bea
C_i &\propto& \frac{\beta(q)}{\alpha(q)} ~=~
-\lim_{u\rightarrow\infty} \frac{y_1}{y_2}
\;,
\eea
which can be calculated numerically for each value of $q$.
A first numerical check suggests that the three ratios
\begin{equation}
C_{(5,1)}\;,\quad C_{(1,20)}\;,\quad C_{(3,6)}\;,
\end{equation}
behave like $q^{6/5}$ for large values of $q$. More precisely, for large $q$, these ratios can be fit by a function 
\begin{equation}
 C_i ~=~ a_i + b_i\, q^{c_i}\,,
\end{equation}
with
\begin{center}
 \begin{tabular}{rrr}
 $a_{(5,1)}=1.72$\,, & $b_{(5,1)}=0.37$\,, & $c_{(5,1)}=1.19$\,, \\
 $a_{(1,20)}=1.29$\,, & $b_{(1,20)}=0.37$\,, & $c_{(1,20)}=1.20$\,, \\
 $a_{(3,6)}=-18.96$\,, & $b_{(3,6)}=0.80$\,, & $c_{(3,6)}=1.20$\,. \\
\end{tabular}
\end{center}
In figure~\ref{num}, we have plotted the normalized ratios
\begin{equation}
 r_i(q) \equiv \frac{1}{b_i} \Big( \frac{\tilde{x}_{i\, (5)} (q)}{\tilde{x}_{i\, (2)}(q)} -a_i \Big)
 \;,
\end{equation}
in log-log scales, and compared them to the power law $q^{6/5}$
of the undeformed BFSS model \eqref{44ex}. Asymptotically in $q$ we find complete agreement, in accordance with our interpretation of the model as a deformation of BFSS.  
\begin{figure}[h!]
 \centering 
\includegraphics[scale=.8]{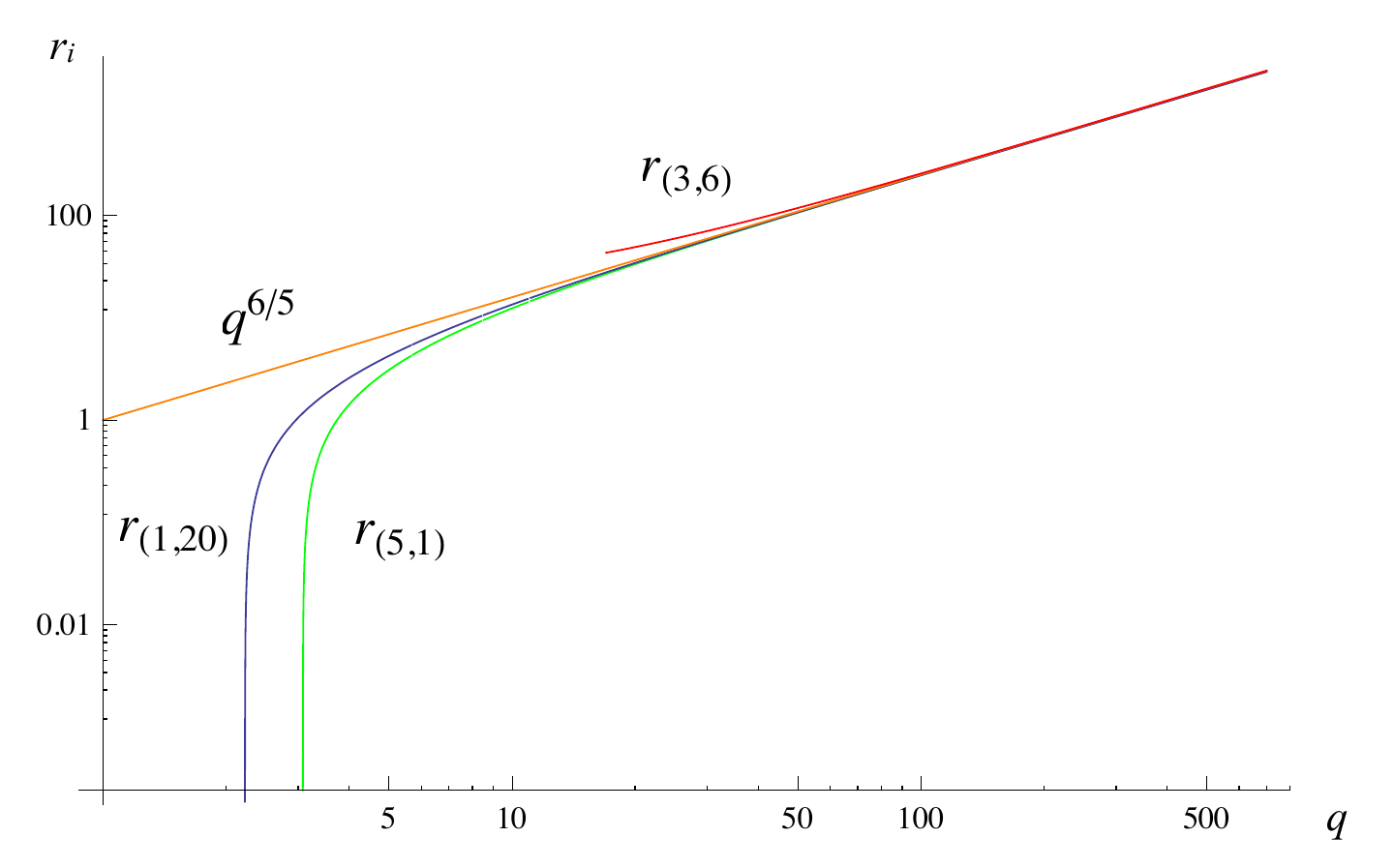}
\caption{{\small Numerical plot of $C_i$ for the operators dual to the scalar fields
(\ref{44bmn}).}}
\label{num}
\end{figure}

\section{Discussion}
\label{sec:discussion}

We have computed two-point scalar correlation functions in 
the strong-coupling regime of the BFSS matrix model. The calculation was performed holographically, using as gravitational dual 
a half-supersymmetric domain wall of the two-dimensional maximally supersymmetric $SO(9)$ gauged supergravity of \cite{Ortiz:2012ib}. 
This two-dimensional domain wall uplifts to a conformal $AdS_2$ times $S^8$ 
geometry which is the near horizon limit of $N$ $D0$ branes; a further uplift to 
eleven dimensions gives an $SO(9)$-symmetric pp-wave. 
Our results are in agreement with those of~\cite{Sekino:1999av,Sekino:2000mg,Hanada:2011fq}.

Furthermore we have constructed a `deformed' half-supersymmetric domain wall 
which uplifts to an eleven-dimensional pp-wave with broken 
$SO(3)\times SO(6)$ symmetry. We have argued that this deformation 
corresponds holographically to turning on an operator vev in the 
matrix model, and we have used the deformed domain wall as gravitational dual in 
order to perform a holographic computation of two-point scalar correlation functions. As a consistency check 
we have verified numerically that 
in the UV-limit all correlators reduce to those computed in the undeformed BFSS matrix model. 
This is in accordance with the fact that in the limit of small radial direction the deformed domain-wall solution asymptotes the undeformed domain wall. 
In principle, similar deformations may exist preserving other maximal subgroups of $SO(9)$.
We have chosen $SO(3)\times SO(6)$ since these correspond to the symmetries of the well known 
BMN operator deformation. However, the corresponding supersymmetric domain wall turned out to be 
related to a vev rather than to an operator deformation of the BFSS matrix model. Indeed, one may expect 
that the geometry dual to the BMN matrix model also requires non-vanishing profiles for the 
axion fields, c.f.~\cite{Lin:2004nb,Lin:2004kw}.

The holographic methods of the present paper can be straightforwardly extended to  compute matrix model $n$-point functions with $n>2$, which could then in principle be checked independently using Monte Carlo methods directly on the matrix quantum mechanics side. 
Another possible direction would be the computation of correlation functions in 
the background of black hole solutions, which corresponds holographically to 
 matrix quantum mechanics at finite temperature.
It would also be very interesting to apply these methods to a background which is holographically
dual to an operator deformation of the BFSS model, such as the BMN matrix model of~\cite{Berenstein:2002jq}.
We plan to return to these questions in the future.

\appendix

\section*{Appendix}

\section{Holographic duals of matrix quantum mechanics}
\label{app:validity}

In this appendix we review, following closely \cite{Polchinski:1999br}, the different holographic dualities of the matrix model and their respective regimes of validity. Matrix theory is obtained from weakly-coupled IIA string theory with $N$ D0 branes in the limit:
\begin{equation}
\label{1r}
g_s\rightarrow0~,~~~l_p=\mathrm{fixed}~,
\end{equation}
where $l_s$ is the string length, $g_s$ is the string coupling and  
$l_p=g_s^{\frac13}l_s$ is the Planck length. 
The near-horizon metric of $N$ D0 branes  is given  in the string frame  by  
\begin{equation}\label{2.11}
\text{ds}_{10}^2 =   \big(\frac{r}{r_0}\big)^{7/2} \text{dt}^2 - \big(\frac{r}{r_0}\big)^{-7/2} (\text{dr}^2 + r^2 \, d\Omega_{8}^2)~,
\end{equation}
 provided we identify $r_0=N^{\frac{1}{7}}l_P$  \cite{Horowitz:1991cd}.  
In particular we have:
\begin{equation}\label{3r}
\frac{R(r)}{l_p}=e^{\frac{2\Phi}{3}}
=N^{\frac12}l_p^{\frac{7}{2}}r^{-\frac72}~,
\end{equation}
where $R(r)$ is the eleven-dimensional circle, $\Phi$ is the dilaton, and we have taken the limit $g_s\rightarrow0$. The $r$-dependent string scale is given 
by
\begin{equation}\label{4r}
l_s(r)\equiv l_pe^{-\frac{\Phi}{3}}~,
\end{equation}
and is obtained by promoting $l_s=g_s^{-\frac13}l_p$ to a local equation 
by replacing $g_s$ by $e^\Phi$. Combining (\ref{3r}), (\ref{4r}) we get
\begin{equation}
\frac{r}{l_s(r)}=N^{\frac14}l_p^{\frac34}r^{-\frac34}~.
\end{equation}
The geometry becomes stringy in the region $r\lesssim l_s(r)$, in which case the $N$ D0 IIA metric cannot be trusted. Hence we must have $r>>l_s(r)$ for the metric to be valid, which leads to the bound $r<<N^{\frac13}l_p$ .
 
A second condition is obtained by the requirement that $R(r)<<l_p$; at distances $R(r)\gtrsim l_p$ the geometry becomes eleven dimensional and the eleven-dimensional uplift must be used instead of the IIA metric. Taking (\ref{3r}) into account this leads to the condition  $r>>N^{\frac17}l_p$.

To summarize, the D0 brane metric of IIA is a valid description in the region\footnote{
We may compare with the regime of validity given in \cite{Hanada:2011fq} by introducing a local  Yang-Mills coupling $g_{YM}^2(r)\equiv e^{\Phi}l^{-3}_s$ which is obtained by replacing $g_s$ by $e^\Phi$ in $g_{YM}^2=g_sl^{-3}_s$. Similarly we define a local 'tHooft coupling $\lambda(r)\equiv g_{YM}^2(r)N$, in terms of which the bound (\ref{bound1}) reads
\begin{equation}\label{bound2}
\lambda(r)^{-\frac13}<<r<<\lambda(r)^{-\frac13}N^{\frac{10}{21}}~.
\end{equation}
This is the same as the bound (1.2) of \cite{Hanada:2011fq} provided we identify $\lambda(r)$, $r$ here with $\lambda$, $|t-t'|$ in \cite{Hanada:2011fq}. 
}
\begin{equation}\label{bound1}
 N^{\frac17}l_p  <<r<<N^{\frac13}l_p  ~.
\end{equation}
Note that we must have $N>>1$ for the inequalities above to make sense.

$\bullet$ The `Maldacena limit' 

The decoupling limit for $N$ D0 branes is given by \cite{Itzhaki:1998dd}:
\begin{equation}\label{5r}
l_s\rightarrow0~,~~~U\equiv\frac{r}{l_s^2}=\mathrm{fixed}~,
~~~g_{YM}^2\equiv\frac{g_s}{l_s^3}
=\mathrm{fixed}~.
\end{equation}
Via the holographic correspondence matrix theory is then dual to the IIA supergravity solution for $N$ D0 branes, provided the latter can be trusted, i.e. provided (\ref{bound1}) holds. 
In order to compare this bound to the corresponding regime of validity given in \cite{Itzhaki:1998dd}, note that at an energy scale $U$ the effective coupling of the Yang-Mills theory is given by
\begin{equation}
g^2_{eff}=g_{YM}^2NU^{-3}~.
\end{equation}
Moreover we have $r=g^{-\frac{2}{3}}_{eff}N^{\frac13}l_p$, as follows from the definitions of $g_{eff}$, $U$; inserting this expression for $r$ in (\ref{bound1}) we obtain
\begin{equation}\label{bound3}
1<<g^2_{eff}<<N^{\frac{4}{7}}~,
\end{equation}
which indeed agrees with \cite{Itzhaki:1998dd}. Note that this implies that N must be large and that the Yang-Mills theory must be strongly coupled in order for IIA supergravity to be a good dual description.

At first sight the limit (\ref{5r}) looks different from (\ref{1r}). However comparing dimensionless quantities, we see that in both cases $g_s\rightarrow0$ and $r/l_p=\mathrm{fixed}$. In either description we have a duality between matrix theory and IIA supergravity with $N$ D0 branes, provided we are in the range given by (\ref{bound1}) or, equivalently, (\ref{bound3}) \cite{Polchinski:1999br}.

$\bullet$ Uplift to eleven dimensions and BFSS

The uplift of the $N$ D0 brane metric of IIA to eleven dimensions gives the metric 
\begin{equation}\label{11d}
ds^2=dx^{+}dx^{-}+\frac{Nl_p^9}{r^7R^{2}}(dx^{-})^2
+ds^2(\mathbb{R}^9)
\end{equation}
with periodicity  $x^+\sim x^++R$, $x^-\sim x^--R$, where $z=x^++x^-$ is the M-theory circle. Performing an infinite boost along the $(t,z)$ directions gives the pp-wave background
\begin{equation}\label{11dboost}
ds^2=dx^{+\prime}dx^{-\prime}+\frac{Nl_p^9}{r^7R^{\prime2}}(dx^{-\prime})^2
+ds^2(\mathbb{R}^9)~,
\end{equation}
in terms of the boosted coordinates $x^{\pm\prime}=t'\pm z'$; $R'$ is the boosted eleventh-dimensional radius, so that the Lorentz boost factor is infinite, $\gamma=R'/R\rightarrow\infty$ with $R'$ fixed. 
Hence the periodic identification now reads: 
$x^{+\prime}\sim x^{+\prime}$, $x^{-\prime}\sim x^{-\prime}-2R'$, i.e. the compactification circle is lightlike.

As already discussed, the description in terms of the eleven-dimensional metric (\ref{11d}) can only be trusted at distances $R(r)>>l_p$, which leads to the condition $r<<N^{\frac17}l_p$. An additional condition comes from the observation that the uplift (\ref{11d}) describes a smeared metric, i.e. one that possesses translational invariance along the eleventh-dimensional circle parameterized by $z$. At distances $r\lesssim R(r)$ this description breaks down, which leads to the condition $r>>N^{\frac19}l_p$. 

To summarize: the lightlike compactification of eleven-dimensional supergravity in the pp-wave background (\ref{11dboost}) is a valid description of matrix theory 
in the region 
\begin{equation}
N^{\frac19}l_p<<r<<N^{\frac17}l_p~.
\end{equation}

\section{Ambiguity $\Delta_{\pm}$}
\label{app:amb}

Consider a KG equation of the form
\begin{equation}\label{1}
\nabla^2 Z-M^2 Z=0~,
\end{equation}
for a bulk AdS$_{d+1}$ scalar field $Z$ dual to a dimension-$\Delta$ operator in 
the boundary CFT. The near-boundary analysis relates $m^2$ to $\Delta$ via
\begin{equation}\label{amb}
\Delta(\Delta-d)=M^2~,
\end{equation}
with $d=1$ in our case.

It is known~\cite{Breitenlohner:1982jf} that for $m^2$ in the range
\begin{equation}\label{range}
-\frac{d^2}{4}<M^2<-\frac{d^2}{4}+1~,
\end{equation}
there are two different AdS-invariant quantizations of the field $Z$, i.e. the 
Lagrangian for $Z$ gives rise to two different quantum theories in AdS. These 
two bulk quantum theories correspond to two different CFT's on the boundary, one for each root of $\Delta$ in (\ref{amb}). Typically one of the dual CFT's will be supersymmetric while the other will be non-supersymmetric~\cite{Klebanov:1999tb}.

For an AdS$_2$ metric (after euclidean rotation to the hyperbolic two-plane) given by
\begin{equation}
ds^2=\frac{1}{r}dt^2+\frac{1}{4r^2}dr^2~,
\end{equation}
it  can be shown that an equation of the form
\begin{equation}
\nabla^{\mu}\left(r^\delta\partial_{\mu}y\right)=-m^2 ~\!r^\delta y~,
\end{equation}
becomes equivalent to (\ref{1}) upon setting
\begin{equation}
y=r^{-\frac{\delta}{2}}Z~,~~~M^2=-m^2+\delta(\delta-1)~.
\end{equation}
We will apply the latter formula to determine $M^2$ in the two cases
corresponding to the scalars in the ${\bf 44}$ and the ${\bf 84}$, respectively.
From (\ref{eomy}), we deduce that
\begin{itemize}
\item the scalar $y_{(44)}$ is obtained for $\delta=-\frac{9}{10}$, $\lambda=-\frac{8}{5}$ which gives $M^2=0.11$\,.
\item The scalar $y_{(84)}$ is obtained for $\delta=-\frac{3}{10}$, $\lambda=-\frac{12}{25}$ which gives $M^2=-0.09$\,.
\end{itemize}
Hence both our examples of scalar fields are in the ambiguous range
and we will need further criteria to determine the dictionary to the boundary theory.


\begin{thebibliography}{10}

\bibitem{Banks:1996vh}
T.~Banks, W.~Fischler, S.~Shenker, and L.~Susskind, {\it {M theory as a matrix
  model: A Conjecture}},  {\em Phys.Rev.} {\bf D55} (1997) 5112--5128,
  [\href{http://xxx.lanl.gov/abs/hep-th/9610043}{{\tt hep-th/9610043}}].

\bibitem{Itzhaki:1998dd}
N.~Itzhaki, J.~M. Maldacena, J.~Sonnenschein, and S.~Yankielowicz, {\it
  Supergravity and the large ${N}$ limit of theories with sixteen
  supercharges},  {\em Phys. Rev.} {\bf D58} (1998) 046004,
  [\href{http://xxx.lanl.gov/abs/hep-th/9802042}{{\tt hep-th/9802042}}].

\bibitem{Boonstra:1998mp}
H.~J. Boonstra, K.~Skenderis, and P.~K. Townsend, {\it The domain wall/{QFT}
  correspondence},  {\em JHEP} {\bf 01} (1999) 003,
  [\href{http://xxx.lanl.gov/abs/hep-th/9807137}{{\tt hep-th/9807137}}].

\bibitem{Kanitscheider:2008kd}
I.~Kanitscheider, K.~Skenderis, and M.~Taylor, {\it {Precision holography for
  non-conformal branes}},  {\em JHEP} {\bf 0809} (2008) 094,
  [\href{http://xxx.lanl.gov/abs/0807.3324}{{\tt 0807.3324}}].

\bibitem{Kanitscheider:2009as}
I.~Kanitscheider and K.~Skenderis, {\it {Universal hydrodynamics of
  non-conformal branes}},  {\em JHEP} {\bf 0904} (2009) 062,
  [\href{http://xxx.lanl.gov/abs/0901.1487}{{\tt 0901.1487}}].

\bibitem{Skenderis:2006uy}  K.~Skenderis and M.~Taylor,
  {\it Kaluza-Klein holography},
  {\em JHEP} {\bf 0605}, 057 (2006)
  [hep-th/0603016].

\bibitem{Bianchi:2001de}
M.~Bianchi, D.~Z. Freedman, and K.~Skenderis, {\it {How to go with an RG
  flow}},  {\em JHEP} {\bf 0108} (2001) 041,
  [\href{http://xxx.lanl.gov/abs/hep-th/0105276}{{\tt hep-th/0105276}}].

\bibitem{Bianchi:2001kw}
M.~Bianchi, D.~Z. Freedman, and K.~Skenderis, {\it {Holographic
  renormalization}},  {\em Nucl. Phys.} {\bf B631} (2002) 159--194,
  [\href{http://xxx.lanl.gov/abs/hep-th/0112119}{{\tt hep-th/0112119}}].

\bibitem{Skenderis:2002wp}
K.~Skenderis, {\it {Lecture notes on holographic renormalization}},  {\em
  Class.Quant.Grav.} {\bf 19} (2002) 5849--5876,
  [\href{http://xxx.lanl.gov/abs/hep-th/0209067}{{\tt hep-th/0209067}}].

\bibitem{Ortiz:2012ib}
T.~Ortiz and H.~Samtleben, {\it {$SO(9)$} supergravity in two dimensions},
  {\em JHEP} {\bf 1301} (2013) 183,
  [\href{http://xxx.lanl.gov/abs/1210.4266}{{\tt 1210.4266}}].

\bibitem{Anabalon:2013zka}
A.~Anabal{\'o}n, T.~Ortiz, and H.~Samtleben, {\it {Rotating D0-branes and
  consistent truncations of supergravity}},  {\em Phys.Lett.} {\bf B727} (2013)
  516--523, [\href{http://xxx.lanl.gov/abs/1310.1321}{{\tt 1310.1321}}].

\bibitem{Sekino:1999av}
Y.~Sekino and T.~Yoneya, {\it Generalized {AdS / CFT} correspondence for matrix
  theory in the large {N} limit},  {\em Nucl.Phys.} {\bf B570} (2000) 174--206,
  [\href{http://xxx.lanl.gov/abs/hep-th/9907029}{{\tt hep-th/9907029}}].

\bibitem{Sekino:2000mg}
Y.~Sekino, {\it Supercurrents in matrix theory and the generalized {AdS / CFT}
  correspondence},  {\em Nucl.Phys.} {\bf B602} (2001) 147--171,
  [\href{http://xxx.lanl.gov/abs/hep-th/0011122}{{\tt hep-th/0011122}}].

\bibitem{Hanada:2011fq}
M.~Hanada, J.~Nishimura, Y.~Sekino, and T.~Yoneya, {\it Direct test of the
  gauge-gravity correspondence for {M}atrix theory correlation functions},
  {\em JHEP} {\bf 1112} (2011) 020,
  [\href{http://xxx.lanl.gov/abs/1108.5153}{{\tt 1108.5153}}].

\bibitem{Berenstein:2002jq}
D.~Berenstein, J.~M. Maldacena, and H.~Nastase, {\it Strings in flat space and
  pp waves from ${N} = 4$ super {Y}ang {M}ills},  {\em JHEP} {\bf 04} (2002)
  013, [\href{http://xxx.lanl.gov/abs/hep-th/0202021}{{\tt hep-th/0202021}}].

\bibitem{KowalskiGlikman:1984wv}
J.~Kowalski-Glikman, {\it Vacuum states in supersymmetric {K}aluza-{K}lein
  theory},  {\em Phys.Lett.} {\bf B134} (1984) 194--196.

\bibitem{figuroa}
M.~Blau, J.~M. Figueroa-O'Farrill, C.~Hull, and G.~Papadopoulos, {\it Penrose
  limits and maximal supersymmetry},  {\em Class. Quant. Grav.} {\bf 19} (2002)
  L87--L95, [\href{http://xxx.lanl.gov/abs/hep-th/0201081}{{\tt
  hep-th/0201081}}].

\bibitem{Gauntlett:2002cs}
J.~P. Gauntlett and C.~M. Hull, {\it {Pp-waves in 11 dimensions with extra
  supersymmetry}},  {\em JHEP} {\bf 0206} (2002) 013,
  [\href{http://xxx.lanl.gov/abs/hep-th/0203255}{{\tt hep-th/0203255}}].

\bibitem{Lin:2004nb}
H.~Lin, O.~Lunin, and J.~M. Maldacena, {\it {Bubbling AdS space and 1/2 BPS
  geometries}},  {\em JHEP} {\bf 0410} (2004) 025,
  [\href{http://xxx.lanl.gov/abs/hep-th/0409174}{{\tt hep-th/0409174}}].

\bibitem{Hull:1984vh}
C.~Hull, {\it {Exact $p p$ Wave Solutions of Eleven-dimensional Supergravity}},
   {\em Phys.Lett.} {\bf B139} (1984) 39.

\bibitem{Townsend:1998qp}
P.~Townsend, {\it The {M}(atrix) model/$ad{S}_2$ correspondence},
  \href{http://xxx.lanl.gov/abs/hep-th/9903043}{{\tt hep-th/9903043}}.

\bibitem{Freedman:1999gk}
  D.~Z.~Freedman, S.~S.~Gubser, K.~Pilch and N.~P.~Warner,
  {\it Continuous distributions of D3-branes and gauged supergravity},
  JHEP {\bf 0007} (2000) 038
  [hep-th/9906194].

\bibitem{Klebanov:1999tb}
I.~R. Klebanov and E.~Witten, {\it {AdS / CFT correspondence and symmetry
  breaking}},  {\em Nucl.Phys.} {\bf B556} (1999) 89--114,
  [\href{http://xxx.lanl.gov/abs/hep-th/9905104}{{\tt hep-th/9905104}}].

\bibitem{deWit:1988ig}
B.~de~Wit, J.~Hoppe, and H.~Nicolai, {\it On the quantum mechanics of
  supermembranes},  {\em Nucl.Phys.} {\bf B305} (1988) 545.

\bibitem{OColgain:2012wv}
E.~O~Colgain, {\it {Beyond LLM in M-theory}},  {\em JHEP} {\bf 1212} (2012)
  023, [\href{http://xxx.lanl.gov/abs/1208.5979}{{\tt 1208.5979}}].

\bibitem{Berg:2001ty}
M.~Berg and H.~Samtleben, {\it An exact holographic {RG} flow between 2d
  conformal fixed points},  {\em JHEP} {\bf 05} (2002) 006,
  [\href{http://xxx.lanl.gov/abs/hep-th/0112154}{{\tt hep-th/0112154}}].

\bibitem{Berg:2002hy}
M.~Berg and H.~Samtleben, {\it Holographic correlators in a flow to a fixed
  point},  {\em JHEP} {\bf 12} (2002) 070,
  [\href{http://xxx.lanl.gov/abs/hep-th/0209191}{{\tt hep-th/0209191}}].

\bibitem{Lin:2004kw}
H.~Lin, {\it {The supergravity dual of the BMN matrix model}},  {\em JHEP} {\bf 0412} (2004) 001,
  [\href{http://xxx.lanl.gov/abs/hep-th/0407250}{{\tt hep-th/0407250}}].

\bibitem{Polchinski:1999br}
J.~Polchinski, {\it {M theory and the light cone}},  {\em
  Prog.Theor.Phys.Suppl.} {\bf 134} (1999) 158--170,
  [\href{http://xxx.lanl.gov/abs/hep-th/9903165}{{\tt hep-th/9903165}}].

\bibitem{Horowitz:1991cd}
G.~T. Horowitz and A.~Strominger, {\it Black strings and $p$-branes},  {\em
  Nucl.Phys.} {\bf B360} (1991) 197--209.

\bibitem{Breitenlohner:1982jf}
P.~Breitenlohner and D.~Z. Freedman, {\it Stability in gauged extended
  supergravity},  {\em Annals Phys.} {\bf 144} (1982) 249.

\end{thebibliography}

\providecommand{\href}[2]{#2}\begingroup\raggedright\endgroup

\end{document}